\documentclass[conference]{IEEEtran}
\ifCLASSINFOpdf
\else
\fi
\usepackage[all]{nowidow}
\usepackage{graphicx}
\usepackage[linesnumbered,lined,vlined,ruled,commentsnumbered,noend]{algorithm2e}
\usepackage{textcomp}
\usepackage{xcolor}
\colorlet{RED}{red}
\usepackage{listings}
\usepackage{url}

\usepackage{xspace}
\usepackage{multirow}
\usepackage{subfloat}
\usepackage{balance}
\usepackage{ctable}
\usepackage{listings}
\usepackage{color}
\usepackage[nobiblatex]{xurl}

\usepackage{multirow}

\usepackage{algpseudocode}
\usepackage{caption}

\usepackage{cleveref}
\usepackage{acronym}
\usepackage{fancybox}
\usepackage{verbatim}
\usepackage[normalem]{ulem}
\usepackage{float}
\usepackage{newfloat}
\usepackage{mdframed}
\let\labelindent\relax
\usepackage{enumitem}
\usepackage{booktabs}
\usepackage{pifont}
\usepackage[ruled,linesnumbered]{algorithm2e}

\usepackage{subfigure}
\usepackage{ulem}
\usepackage[flushleft]{threeparttable}
\usepackage{booktabs}
\usepackage{tabularray}

\definecolor{red}{RGB}{255,0,0}
\definecolor{green}{RGB}{18,220,168}

\newcommand{\ie}{i.e.,\xspace}

\newcommand{\attcandi}{terminals\xspace}

\newcommand{\graphacc}{graph-level accuracy\xspace}
\newcommand{\graphprecision}{graph-level precision\xspace}
\newcommand{\graphrecall}{graph-level recall\xspace}

\newcommand{\nodeacc}{node-level accuracy\xspace}
\newcommand{\nodeprecision}{node-level precision\xspace}
\newcommand{\noderecall}{node-level recall\xspace}

\usepackage{pifont}

\newcommand{\del}[1]{\textcolor{red}{}}

\newcommand{\toolname}{\textsc{NodLink}\xspace}

\newcommand{\insighttwo}{attack polymerism\xspace}
\newcommand{\anosec}{Sangfor\xspace}

\newcommand{\eat}[1]{}

\acrodef{wp}[WP]{Website Fingerprinting}
\acrodef{apt}[APT]{Advanced Persistent Threats}
\acrodef{lol}[LotL]{Living-Off-The-Land}
\acrodef{ids}[IDS]{Intrusion Detection System}
\acrodef{vae}[VAE]{Variational AutoEncoder}
\acrodef{re}[RE]{reconstruction error}
\acrodef{sv}[SV]{Stableness Value}
\acrodef{as}[AS]{anomaly score}
\acrodef{has}[HAS]{hopset anomaly score}
\acrodef{etw}[ETW]{Event Tracing for Windows}
\acrodef{e3}[E3]{Engagement 3}
\acrodef{e5}[E5]{Engagement 5}
\acrodef{ttp}[TTPs]{Tactics, Techniques, and Procedures}
\acrodef{hsg}[HSG]{High-level Scenario Graph}
\acrodef{nlp}[NLP]{Natural Language Processing}
\acrodef{dg}[DG]{Detection Graph}
\acrodef{poi}[POI]{Point of Interest}
\acrodef{iv}[IV]{Important Value}
\acrodef{ioc}[IOC]{Indicators of Compromise}
\acrodef{ag}[AG]{Attack Graph}
\acrodef{cs}[CS]{Cobalt Strike}
\acrodef{mttd}[MTTD]{Mean Time to Detect}
\acrodef{soc}[SOC]{Security Operations Center}
\acrodef{loc}[LoC]{Lines of Code}
\acrodef{wmi}[WMI]{Windows Management Instrumentation}
\acrodef{dlls}[DLLs]{Dynamic-Link Libraries}
\acrodef{sota}[SOTA]{state-of-the-art}
\acrodef{edr}[EDR]{Endpoint Detection and Response}
\acrodef{nlp}[NLP]{Natural Language Processing}
\acrodef{stp}[STP]{Steiner Tree Problem}
\acrodef{isg}[ISG]{Importance-Score-Guided Search}

\newlength{\MaxSizeOfLineNumbers}%
\definecolor{keywordcolor}{rgb}{0.8,0.1,0.5}
\definecolor{lightlightgray}{gray}{.96}
\definecolor{lightgray}{gray}{.925}
\definecolor{medlightgray}{gray}{0.7}
\definecolor{medgray}{gray}{0.4}
\definecolor{darkgray}{gray}{0.35}
\definecolor{nearblack}{gray}{0.15}

\crefname{component}{Component}{Components}

\newcommand{\distance}{8pt}
\begin{document}
%
\title{\toolname: An Online System for Fine-Grained APT Attack Detection and Investigation}


%
\author{\IEEEauthorblockN{Shaofei Li\IEEEauthorrefmark{2},
Feng Dong\IEEEauthorrefmark{3},
Xusheng Xiao\IEEEauthorrefmark{4},
Haoyu Wang\IEEEauthorrefmark{3},
Fei Shao\IEEEauthorrefmark{5},
Jiedong Chen\IEEEauthorrefmark{6}, Yao Guo\IEEEauthorrefmark{2},\\  Xiangqun Chen\IEEEauthorrefmark{2}, and Ding Li\thanks{* is the corresponding author.}\IEEEauthorrefmark{2}\IEEEauthorrefmark{1}}
\IEEEauthorblockA{\IEEEauthorrefmark{2}Key Laboratory of High-Confidence Software Technologies (MOE), School of Computer Science, Peking University}
\IEEEauthorblockA{\IEEEauthorrefmark{3}Huazhong University of Science and Technology\IEEEauthorrefmark{7}\thanks{\IEEEauthorrefmark{7} Hubei Key Laboratory of Distributed System Security, Hubei Engineering Research Center on Big Data Security, School of Cyber Science and Engineering, Huazhong University of Science and Technology.}, \IEEEauthorrefmark{4}Arizona State University}
\IEEEauthorblockA{\IEEEauthorrefmark{5}Case Western Reserve University, \IEEEauthorrefmark{6}Sangfor Technologies Inc.}
\IEEEauthorblockA{\IEEEauthorrefmark{2}\{lishaofei, ding\_li, yaoguo, cherry\}@pku.edu.cn, \IEEEauthorrefmark{3}\{dongfeng, haoyuwang\}@hust.edu.cn}
\IEEEauthorblockA{\IEEEauthorrefmark{4}xusheng.xiao@asu.edu, \IEEEauthorrefmark{5}fxs128@case.edu, \IEEEauthorrefmark{6}chenjiedong1027@gmail.com}}


\IEEEoverridecommandlockouts
\makeatletter\def\@IEEEpubidpullup{6.5\baselineskip}\makeatother
\IEEEpubid{\parbox{\columnwidth}{
    Network and Distributed System Security (NDSS) Symposium 2024\\
    26 February - 1 March 2024, San Diego, CA, USA\\
    ISBN 1-891562-93-2\\
    https://dx.doi.org/10.14722/ndss.2024.23204\\
    www.ndss-symposium.org\\
}
\hspace{\columnsep}\makebox[\columnwidth]{}}

\maketitle

\begin{abstract}
\ac{apt} attacks have plagued modern enterprises, causing significant financial losses.
To counter these attacks, researchers propose techniques that capture the complex and stealthy scenarios of APT attacks by using provenance graphs to model system entities and their dependencies. Particularly, to accelerate attack detection and reduce financial losses, online provenance-based detection systems that detect and investigate APT attacks under the constraints of timeliness and limited resources are in dire need.
Unfortunately, existing online systems usually sacrifice detection granularity to reduce computational complexity and produce provenance graphs with more than 100,000 nodes, posing challenges for security admins to interpret the detection results. 
In this paper, we design and implement \toolname, the first online detection system that maintains high detection accuracy without sacrificing detection granularity. Our insight is that the \ac{apt} attack detection process in online provenance-based detection systems can be modeled as a \acf{stp}, which has efficient online approximation algorithms that recover concise attack-related provenance graphs with a theoretically bounded error. To utilize the frameworks of the \ac{stp} approximation algorithm for \ac{apt} attack detection, we propose a novel design of in-memory cache, an efficient attack screening method, and a new \ac{stp} approximation algorithm that is more efficient than the conventional one in \ac{apt} attack detection while maintaining the same complexity.
We evaluate \toolname in a \emph{production environment}. The open-world experiment shows that \toolname outperforms two \ac{sota} online provenance analysis systems by achieving magnitudes higher detection and investigation accuracy while having the same or higher throughput.
\end{abstract}



\section{Introduction}
Advanced Persistent Threat (\ac{apt}) attacks have become a major threat to modern enterprises~\cite{apttrends,targeted}. Existing \ac{edr} systems adopted by these enterprises to defend against cyber attacks have difficulties in countering \ac{apt} attacks due to the lack of capability to recover the complex causality relationships between the steps of \ac{apt} attacks~\cite{6620049,idika2007survey,10.1145/1127345.1127348,10.1145/3319535.3363224,lifirst,10.1145/3422337.3447833}. Therefore, practitioners and researchers~\cite{10.1145/3319535.3363217,263852,277080,king2003backtracking,215975,216025,10.5555/1855807.1855817} now analyze the system auditing events in provenance data to recover \ac{apt} attack scenarios. Unfortunately, most of the existing provenance analysis systems only support post-mortem analysis for alerts of \ac{edr} systems, which can delay the accurate detection of \ac{apt} attacks for a week~\cite{liu2018towards} and cause significant financial losses. As shown in a recent study, it costs an enterprise about \$32,000 each day when an attacker persists in the network~\cite{Incident-Investigation}.

To this end, researchers have built online provenance-based detection systems that detect and investigate \ac{apt} attacks simultaneously~\cite{sleuth, HOLMES, Han2020,9152772,9152771}. Unlike post-mortem analysis systems, online provenance-based detection systems can detect and recover the logic of an \ac{apt} attack within seconds of its occurrence, allowing security admins to respond in time and reduce potential losses. Furthermore, by conducting a comprehensive analysis on the whole \ac{apt} attack campaign instead of individual system events, online provenance-based detection systems have substantially fewer false positives than conventional \ac{edr} systems, further improving the effectiveness and efficiency of \ac{apt} attack investigation~\cite{hassan2019nodoze,10.1145/3427228.3427255}.

Despite these promising early results, building an accurate online detection system is still conceptually challenging due to \textit{the constraint of limited resources} and \textit{the high expectation of timeliness.} \del{\sout{A recent report has shown that enterprises only spend 0.3\% of their revenue on security on average~\cite{secspend}.}}
Recent research has shown that the operating cost is the primary bottleneck for the industry to adopt an EDR system~\cite{dong2023yet}.
Thus, it is important to reduce the running cost of provenance-based detection system. Meanwhile, people still expect a provenance-based detection system to detect \ac{apt} attacks in a timely online manner. Unfortunately, achieving high detection accuracy under the constraints of timeliness and limited resources is particularly challenging as provenance data is a highly structured graph (called provenance graph)~\cite{hassan2019nodoze,wang2020you}. 
Existing graph processing algorithms,  such as graph neural networks~\cite{han2021sigl} or iterative message passing algorithms~\cite{hassan2019nodoze}, cannot be directly applied due to their low efficiency. 

To address these challenges, recent online provenance-based detection systems over-approximate the highly structured provenance graph with low-dimensional data structures~\cite{Han2020} or manually crafted rules~\cite{sleuth, HOLMES,9152772,9152771}.
While these systems reduce the computational complexity by sacrificing detection granularity, they make the detection result hard to interpret. 
For instance, given an \ac{apt} attack captured in the provenance data, one of the \ac{sota} online detection systems, UNICORN~\cite{Han2020}, may generate a provenance graph with more than 100,000 nodes. However, only fewer than 100 nodes are related to the \ac{apt} attack. 
Thus, identifying the attack steps represented by such a small number of nodes out of the provenance graph is like ``searching for a needle in a haystack,'' which is extremely difficult for the security admins.

Generally, the workflow of provenance-based \ac{apt} detection~\cite{sleuth, HOLMES,9152772,9152771,wang2020you,263852,8450016,han2021sigl,satvat2021extractor} contains three steps: \ding{172} \textit{attack candidate detection}, which selects \ac{ioc} or anomalous events; \ding{173} \textit{provenance graph construction}, which builds graphs with nodes representing system entities and edges representing entity interactions 
 (e.g., read/write files) to uncover the interactions between the \ac{ioc}s or anomalies; \ding{174} \textit{comprehensive detection}, which detects APT attacks based on the provenance graphs built in step \ding{173}. Among the three steps, the quality of the provenance graph built in step \ding{173} is the key to accurate and fine-grained detection. Unfortunately, building an optimally concise and accurate provenance graph that discloses the interactions between the \ac{ioc}s or anomalies is challenging~\cite{277080,hassan2019nodoze}. The heuristics used by existing approaches either overly approximate the provenance graph~\cite{8935406}, miss critical information~\cite{Han2020}, or are too heavy for online detection systems~\cite{wang2020you,han2021sigl}. 
Therefore, they cannot achieve conciseness, efficiency, and accuracy at the same time.

In this paper, we propose the FIRST online detection system that achieves fine-grained detection while maintaining detection accuracy under the constraints of timeliness and limited resources. 
Existing approaches fail to achieve conciseness, efficiency, and accuracy for \ac{apt} detection simultaneously due to their fundamental limitation: lack of formal models of multiple goals of \ac{apt} detection.
Thus, to address this fundamental limitation, we propose to model the provenance graph construction (step \ding{173} of provenance detection) as an \ac{stp} (Steiner Tree Problem)~\cite{hwang1992steiner,Imase1991DynamicST}, which is effective in modeling multiple goals and has efficient online approximation solutions with theoretical bounded errors. 
Consider \ac{ioc}s or anomalies as a set of predefined nodes in \ac{stp} (\ie terminals), and assign each interaction among system entities with the same non-negative weight. 
Then building a provenance graph can be modeled as an online \ac{stp}, which searches for a subgraph that links all anomalies with minimal numbers of edges. 
By doing so, we can design an approximation algorithm that ensures a subgraph with minimal edges within a theoretically bounded error range in polynomial time.

Although it sounds promising, solving the problem of \ac{apt} attack detection based on \ac{stp} faces three main challenges.
The first one is how to detect long-term attacks. \ac{stp} requires knowing the whole provenance graph in advance. However,  keeping all provenance data in memory is impossible due to data size and storing it in an on-disk database is not practical either due to the I/O bottleneck. A straightforward approach of using a time window that only holds the most recent data is not always effective as attackers can take longer than the window allows. To solve this problem, we propose a novel in-memory cache design with a scoring method to prioritize events that may cause APT attacks and capture long-running attacks within the time window of STP. The second challenge is how to efficiently identify terminals in \ac{stp}. Existing detection methods depend on intensive random walking and message passing on the provenance graph~\cite{han2021sigl,hassan2019nodoze,wang2020you}, which is not suitable for an online system. To solve this problem, we design an IDF-weighted three-layered \ac{vae} that requires minimal computation.
The last challenge is that the current approximation algorithms for \ac{stp} are still not efficient enough for \ac{apt} attack detection. Existing approaches~\cite{Imase1991DynamicST,xu2022learning,naor2011online,10.1145/2488608.2488674} require finding the shortest path between two nodes, which is too expensive for online \ac{apt} attack detection.  To solve this problem, we develop an importance-oriented greedy algorithm for online \ac{stp} optimization that achieves low computing complexity with a bounded competitive ratio.

 We implement \toolname as a working system and deploy it to a beta version of the commercial \ac{soc} of a security company, \anosec. We evaluate \toolname with an \textit{open-world setting} in the production environments of customers of \anosec, including hospitals, universities, and factories. \textit{This is the \textbf{first} open-world evaluation on provenance-based \ac{apt} detection systems}. During two-day testing, \toolname successfully detected seven \textbf{real} \textbf{attacks} that happened in the customer's production environment. It outperforms the \ac{sota} \ac{apt} attack detection systems, HOLMES~\cite{HOLMES} and UNICORN~\cite{Han2020}, in terms of the accuracy of detecting attacks from the provenance data and the accuracy of revealing attack steps from the detected attacks.
 Particularly, HOLMES fails to detect any attacks due to its incomplete rule set while UNICORN generates seven times more false positives on the graph level and three orders of MAGNITUDE more false positives on the node level. Besides the open-world experiment, we also evaluate \toolname with public datasets and in-lab-made datasets that simulate the internal environments of \anosec. The conclusion is also consistent with the open-world experiments: \toolname consistently outperforms HOLMES and UNICORN by generating magnitudes fewer false positives.

In summary, our main contribution is proposing a formal and rigorous mathematical framework based on STP to model APT detection. This model enables efficient and accurate detection of APT attacks with valid approximation bounds. This is the first paper with theoretical analysis of APT detection error, to our best knowledge.
 Further, we did not simply apply the STP model since it is not efficient enough for online APT detection (See Section V.C HopSet Construction). Instead, we propose a novel SPT framework that is more efficient while maintaining the approximation error. We also provide new theoretical analysis of the error of our HopSet Construction in Section V.E. See Section V.C for why STP is not efficient enough.
 We summarize our major contributions as follows:
\begin{itemize}[noitemsep, topsep=1pt, partopsep=1pt, listparindent=\parindent, leftmargin=*]
    \item We model the \ac{apt} detection as the online \ac{stp}, which provides a new vision in online \ac{apt} detection. 
    \item We design and implement an online \ac{apt} detection system, \toolname, that achieves fine-grained detection with timeliness and limited resources based on STP.   
    \item We evaluate \toolname in real production environments. To the best of our knowledge, this is the first open-world evaluation of provenance-based \ac{apt} attack detection. 
    \item We release an open-source version of \toolname along with a new public provenance dataset for attack detection that simulates the internal environment of \anosec at \url{https://github.com/Nodlink/Simulated-Data}.  
\end{itemize}

\section{Background}
\label{sec:motiv}
\begin{figure*}[t!]
    \centering\includegraphics[width=\textwidth]{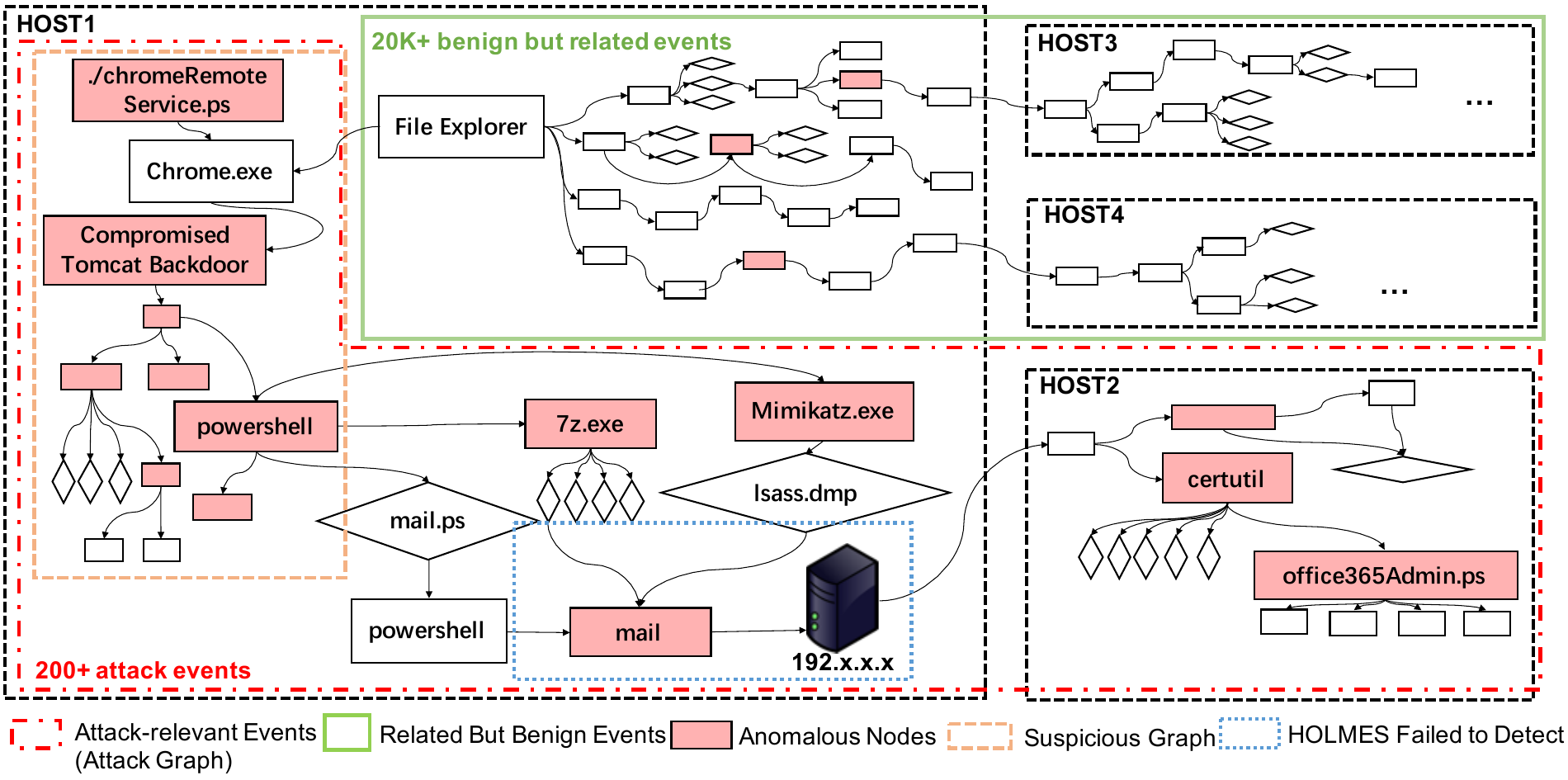}
	\caption{Example provenance graph of APT29. \toolname is able to pinpoint the attack in a concise alert provenance graph with about 200 nodes. \textit{The main part of the alert provenance graph is marked by the red dotted line. Critical attack steps are highlighted in red-shaded rectangles. } }
	\label{fig:motiv}
\end{figure*}

In this section, we introduce the background of provenance analysis and provenance-based detection systems, and describe the metrics to measure their performance.

\subsection{Provenance Analysis}
Provenance analysis systems collect system auditing events of system calls from kernels using system monitoring tools, such as Sysdig~\cite{sysdig} and Linux Audit~\cite{auditd},
and build a provenance graph based on the collected events to show activities between system entities.
A provenance graph is a directed graph. Its nodes are system entities, such as processes, files, and IP addresses. The edges of a provenance graph are control and data flows between system entities. For instance, there is an edge from process $p_1$ to process $p_2$ if $p_1$ forks $p_2$. Similarly, an edge links process $p_1$ and file $f_1$ if $p_1$ writes data to $f_1$. 
Currently, several pieces of research are developed to build and analyze provenance graphs. These approaches span from system monitoring tools~\cite{karande2017sgx,paccagnella2020logging}, data storage~\cite{fei2021seal,217579,10.1145/2508859.2516731,10.1145/3243734.3243763,10.1145/2976749.2978378}, and attack detection and investigation~\cite{sleuth,HOLMES,9152772,Han2020,10.1145/3319535.3363217,263852}. 
Following the existing work~\cite{277080,wang2020you,HOLMES,9152772,sleuth,263852,zeng2021watson},
we focus on the system events that are critical to attack steps, which we list in Table~\ref{tab:system-call}.

Despite the effectiveness of provenance analysis systems, their data logging can generate a colossal amount of records, which introduces significant pressure in log processing, and thus most existing provenance analysis systems only support post-mortem analysis for alerts of \ac{edr} systems.
To address this problem, provenance-based detection system has received extensive research in recent years~\cite{sleuth,HOLMES,9152772,DBLP:journals/corr/ManzoorMVA16,Han2020,wang2020you,zeng2021watson,SoK-History}.
It takes the \textit{system auditing events} as input and outputs alerts in the form of provenance graphs when it detects attacks. Compared with the conventional \ac{edr} systems that process each event individually, provenance-based detection systems leverage the structure and dependency information of the provenance graph to design detection algorithms. However, existing solutions still suffer from challenges and limitations, as we will discuss below.

\begin{table}[!t]
    \centering
    \caption{System events of provenance analysis}
    \resizebox{0.49\textwidth}{!}{\begin{tabular}{c c}
    \hline
    \textbf{Events} & \textbf{Operation Types} \\ \hline
    Process$\leftrightarrow$File & read, write, create, chmod, rename\\
    Process$\leftrightarrow$Process & fork, clone, execve, pipe \\
    Process$\leftrightarrow$IP & sendto, recvfrom, recvmsg, sendmsg \\
    \hline
    \end{tabular}}

    \label{tab:system-call}
\end{table}

We illustrate the challenges and limitations in detecting and investigating a real-world APT attack (APT29~\cite{apt29}).  
Figure~\ref{fig:motiv} shows the provenance graph of APT29. In APT29, an attacker compromises the Chrome browser on the victim's machine and leverages PowerShell to run malicious scripts. Then the malicious PowerShell script tries to obtain the root privilege on the victim's machine and, at the same time, executes the Mimikatz tool to get user credentials for lateral movement. The attacker then further attacks another host in the same network once he has the credentials of a different user.

\noindent\textbf{Challenges:} The challenges come from the high volume of data and the imbalanced ratio between attack and benign events. For instance, the provenance graph that contains the APT29 in Figure~\ref{fig:motiv} has more than 20K events. However, there are only about 200 events that are \textit{attack-relevant} (shown using red dash-dotted boxes) and more than 99\% of the events are \textit{attack-irrelevant} events introduced by benign system activities (shown using green solid boxes)~\cite{hassan2019nodoze,liu2018towards}. Furthermore, benign events may look similar to attack-related events in the provenance data, which further increases the difficulty in detecting attacks precisely. For instance, in Figure~\ref{fig:motiv}, the attacker leverages PowerShell to run malicious scripts. However, the system administrator may also use PowerShell to maintain the system. Simple solutions, such as a blacklist of process names, do not work because they cannot distinguish whether the attacker or the system administrator uses PowerShell. 

\noindent\textbf{Limitations of Existing Techniques:}
Existing provenance-based detection systems can be categorized into two groups: rule-based and learning-based systems. Numerous instances within both groups have demonstrated the capability to achieve practical \graphacc~\cite{sleuth,HOLMES,9152772,Han2020,10.1145/3319535.3363217,263852,10.1145/3427228.3427255}. 
However, existing systems cannot achieve sufficient \nodeprecision and \noderecall simultaneously. 

Rule-based systems suffer low \nodeacc due to the incomplete rule set. For instance, one of the \ac{sota} rule-based systems, HOLMES~\cite{HOLMES}, cannot detect the attack steps in the blue dotted boxes in Figure~\ref{fig:motiv}. 
The reason is that the rules of HOLMES are initiated from external untrusted IPs, while the \textsf{mail} process does not have a direct connection to the external IP addresses. Thus, the output graph of HOLMES for the attack in Figure~\ref{fig:motiv} is fragmented. It fails to link the attack steps on HOST2 to the attack steps on HOST1, posing challenges for root cause analysis. Besides, rule-based systems generate many false positives on the node level since the rules cannot model all features of a dynamic system well.

On the flip side, learning-based systems have low \nodeprecision due to over-approximation. To support online detection, existing learning-based approaches project the provenance graph into low-dimensional data structures to reduce the computational complexity, leading to over-approximation. For instance, the \ac{sota} learning-based online detection approach, UNICORN~\cite{Han2020}, converts the provenance graph into a hashing vector. Thus, UNICORN cannot pinpoint the attack-relevant nodes in Figure~\ref{fig:motiv} (shown in red dash-dotted boxes) from the benign data (shown in green dash-dotted boxes), leading to a low \nodeprecision ($<$1\%).  

\subsection{Online Steiner Tree Problem}
Online \ac{stp} is a combinatorial optimization problem~\cite{Imase1991DynamicST} determined to be an NP-Complete problem with bounded approximation~\cite{karp2010reducibility,doi:10.1137/0132072,kou1981fast,takahashi1980approximate,mehlhorn1988faster,byrka2010improved,10.5555/645929.672717,doi:10.1137/S0097539793242618}. Given an undirected graph $G = (V,E)$ with non-negative edges weights $w_{e}$ for each edge $e \in E$ and a sequence of online revealed vertices (called terminals) $T = \{t_1, t_2, \dots, t_k\}$, it outputs a subgraph $S_i$ of $G$ that spans $\{t_1, t_2, \dots, t_i\}$. The objective is to minimize the total cost of $c(\bigcup_{i=1}^k S_i)$.

The general framework of online \ac{stp} optimization algorithm is presented in Algorithm~\ref{alg:onlinegreedy}. For each new terminal $t_{i}$ that arrives, we select the set $S_{i}$ of unselected edges that connects $t_{i}$ to the current solution of the algorithm. This set $S_{i}$ is chosen to be the cheapest, which has the minimal weighted edges, among all possible sets that connect $t_{i}$ to the existing terminals. To find this set, it uses a greedy approach: it computes the shortest path $P_{j}$ between $t_{i}$ and each previous terminal $t_{j}$ where $j < i$, and then it picks the cheapest one as $S_{i}$. We add $S_{i}$ to the solution and repeat this process until all terminals have arrived. The final solution is the union of all sets $S_{i}$.

\noindent\textbf{Competitive Ratio: }The competitive ratio is a common measure for evaluating online \ac{stp}. It is the worst-case cost of the online algorithm's solution, denoted by $c(\cup^{k}_{i}S_{i})$, with the cost of an optimal solution that has full knowledge of the input, denoted by $c(O^{*})$. Formally, the competitive ratio is defined as $\frac{c(\cup^{k}_{i}S_{i})}{c(O^{*})}$~\cite{borodin2005online}.

Algorithm~\ref{alg:onlinegreedy} can achieve a competitive ratio of $O(\log(k))$~\cite{Imase1991DynamicST}, where $k$ is the number of terminals. According to theoretical analysis, the competitive ratio of any online Steiner tree algorithm is $O(\log(k))$~\cite{Imase1991DynamicST}. Therefore, the simple greedy algorithm is optimal for online Steiner Tree building.

\RestyleAlgo{ruled}
\begin{algorithm}[!t]

\SetKwInOut{Input}{Input}
\SetKwInOut{Output}{Output}

\caption{Greedy Algorithm for Online \ac{stp}}
\label{alg:onlinegreedy}

\Input{the Undirected Weighted Graph $G$, Terminal Stream $TS$}
\Output{Selected Edge Set $S$}
$T \leftarrow \emptyset$ \\
$S \leftarrow \emptyset$ \\
\While{$t_{i}$ arrives from $TS$}{
    $PATH \leftarrow \emptyset$\\
    \For{$t_{j} \in T$}{
        $P_{j} \leftarrow shortest\_path(t_{i}, t_{j}, G)$\\
        $PATH \leftarrow PATH.add(P_{j})$\\
    }
    $S_{i} \leftarrow get\_min\_cost(PATH)$\\
    $S \leftarrow S \cup S_{i}$\\
    $T \leftarrow T.add(t_{i})$
}
\Return{$S$}

\end{algorithm}

\section{Threat Model} Our threat model is similar to the previous work on learning-based provenance-based attack detection~\cite{hassan2019nodoze,Han2020}. We assume attacks have distinct features, which can be detected by statistical patterns or manually created rules.  We assume that the system-level auditing frameworks are secure, which means that they can faithfully record system activities with sufficient details. In addition, we assume the storage and transmission of system logs are secure. Although there are several attack vectors that may compromise the collection, storage, and transmission of system logs~\cite{michael2020forensic,paccagnella2020logging}, they are beyond the scope of this work. We do not consider the attacks performed using implicit flows (e.g., side channels)  that do not go through kernel-layer auditing and thus cannot be captured by the underlying provenance tracker. Although we allow attacks in the training data, we assume that the majority of the training data are not contaminated by attacks.

\section{Overview of \toolname}\label{sec:overview}

Generally, \toolname is an online APT attack detection system. It accepts an event stream of system provenance events collected from the agents installed on the monitored hosts. The outputs of \toolname are concise alert provenance graphs that contain critical attack steps. An example of alert provenance graph is shown in Figure 1 (marked by the red dotted box
with about 200 events). 

Algorithm~\ref{alg:AlgorithmSTP} shows the high-level workflow of \toolname, which is similar to existing detection systems~\cite{sleuth,HOLMES,9152772,DBLP:journals/corr/ManzoorMVA16,Han2020,wang2020you,zeng2021watson,SoK-History}. Our \toolname algorithm detects anomalies every $\Delta$ ($\Delta$ = 10 in our implementation) seconds through four phases: (1) In Memory Cache Building(line 6), (2) Terminal Identification(line 7), (3) Hopset Construction(line 8), and (4) Comprehensive Detection(line 9). We fetch events and store them in a cache, allowing \toolname to track the causalities of events of long-running attacks in memory without suffering from the performance bottleneck introduced by slow I/O. Suspicious processes are identified and assigned anomaly scores based on local features, such as the command lines, process names, and accessed files. Hopsets are constructed for each terminal to connect topologically close event-level anomalies. Finally, we merge hopsets with the cache and report any subgraph with deviating anomaly scores as an alert to reduce false positives.

The key novelty of our approach is in Phase 3 of Hopset Construction. Existing approaches either use heavy graph learning algorithms~\cite{wang2020you,zengy2022shadewatcher} or heuristics that are error prone~\cite{HOLMES}. In our approach, we propose an online-STP-based approach that ensures conciseness, accuracy, and efficiency. To model the APT attack detection on provenance graph as \ac{stp}, we regard the attack-related processes as terminals and convert the directed graph to an undirected graph with equal non-negative weight $w$ of each edge, $w = 1$ for simplification. It helps us to model terminals that do not have a direct causal relationship but have dependencies on the same node. Thus the object of \ac{apt} detection is to identify the terminal set $T$ first and find the edge set $S$ that can connect all the attack-related nodes with the minimal total weight $c(S) = \sum_{(u,v)\in S}w_{uv}$, which is a standard \ac{stp} formula. Then, our system raises an alert if the aggregated anomaly score of the nodes in the Steiner tree is larger than a threshold.

In addition, we also design a novel in-memory cache that addresses long-running attacks while ensuring efficiency. For online \ac{apt} detection, a time window is required to buffer newly coming events. However, the attacker may abuse a naive time window by taking long-running attacks. To this end, we design the time window as an in-memory cache that keeps updated and anomalous nodes in memory and evicts other nodes to the disk(line 4). While building the Steiner Trees, once our system encounters a node evicted to the disk, it loads it back to the memory. This design ensures that \toolname can handle long-running attacks while maintaining the performance of the online algorithm.

We have also adapted the standard online STP algorithm for APT attack detection. In Algorithm~\ref{alg:onlinegreedy}, the step of finding the shortest path between the newly arrived terminal and having seen terminals at line 6 is too time-consuming for an online detection system. Therefore we design an importance-oriented greedy algorithm that reduces the time complexity from $O(N^2)$ to $O(N)$, where $N$ is the number of nodes, for the step of Hopset Construction. We further provide the theoretical analysis in Section~\ref{TheoreticalAnalysis}.

\RestyleAlgo{ruled}
\begin{algorithm}[!t]

\SetKwInOut{Input}{Input}
\SetKwInOut{Output}{Output}

\caption{Algorithm of \toolname Solving the APT Detection as \ac{stp}}
\label{alg:AlgorithmSTP}

\Input{Event Stream $E$, Time Window $\Delta$}
\Output{Attack Graph $AG$}
$C \leftarrow \emptyset$ \\
$time \leftarrow 0$\\
$E_{t} \leftarrow \emptyset$\\
\While{catching $e$ from $E$}{
    \eIf{$time \ge \Delta$}{
        $G \leftarrow cache\_building(E_{t}$)\\
        $T \leftarrow terminal\_identification(G)$\\
        $C \leftarrow hopset\_construction(G,T)$\\
        $AG \leftarrow comprehensive\_detection(C)$\\
        \If{$AG \neq \emptyset$ }{
            \textbf{Raise Alerts $AG$}
        }
        $E_{t} \leftarrow \emptyset$\\
        $time \leftarrow 0$
    }
    {
        $E_{t}.add(e)$\\
        $time \leftarrow time.increase()$\\

    }
}

\end{algorithm}

\section{Design Details}
In this section, we present the design details of the components of \toolname.

\subsection{In-memory Cache}\label{InmemoryCache}

We design the time window of \toolname as an in-memory cache to address the long-running attacks and maintain the efficiency of the online  \ac{stp} optimization algorithm. It maintains the latest anomalous nodes in memory and evicts the outdated and benign nodes to disk.

 In our design, the in-memory cache contains the edges of the provenance graph in the form of $<srcid, dstid, attr>$, where $srcid$ and $dstid$ are the IDs of the source and destination of an edge, respectively, and $attr$ is the attribute of the edge, which includes the operation types and time stamps. The types of edges and their available operation types are listed in Table~\ref{tab:system-call}. \toolname also stores necessary attributes for nodes in the directed graph, which are processes, files, and IP addresses. For processes, \toolname stores their command lines, process names, pids, and uids. For files and IPs, \toolname stores their paths and IPs with ports, respectively. 

\textit{Cache Updates: } 
Generally, the in-memory cache buffers subgraphs that (1) have higher anomaly scores and (2) are actively evolving. The in-memory cache updates every time window with length $\Delta$ with the solutions of \ac{stp} in the current time window. \toolname utilizes the in-memory cache for global \ac{stp} solution and false positive reduction, which we will discuss in Section~\ref{ComprehensiveDetection} in detail.

\toolname organizes the subgraphs in the cache as hopset and gives each hopset a \ac{has}. On the high level, a hopset is a subset of the provenance graph that contains the local context information of a set of event-level anomalies. \ac{has} indicates the degree of abnormality for each hopset. We will discuss the details of hopset and \ac{has} in the following sections.



%

We define the energy of a hopset, $h$,  as $E=\epsilon^{age}*has(h)$, where $\epsilon$ is a decaying factor, $age$ is the number of time windows passed since the last update to the hopset. $age=0$ if the given hopset has been updated in the last time window (e.g., new events are added to the graph). Otherwise, \toolname increases $age$ by one for every time window passes. When the cache is full, \toolname evicts the hopset with the lowest energy to the Neo4j~\cite{neo4j} database. 

\textit{Retrieving Nodes From the Disk:} To detect long-term attacks, \toolname designs the storage policy of the graph database and retrieves the subgraphs from the database when encountering the evicted nodes. When evicting the hopset to the disk, \toolname stores all the relationships and attributes of nodes and edges, including the anomaly score, and removes them out. \toolname assigns a unique \textit{uuid} for each node using md5 value. For processes, \toolname calculates the md5 of \textit{pid + tid + command line}. For files, \toolname calculates the md5 of \textit{/full/path/filename}. For IPs, \toolname calculates the md5 of \textit{src\_ip:port:dst\_ip:port}.
Thus \toolname can retrieve the attributes of the evicted node and hopset. 

Since each node has a unique \textit{uuid}, we can check if the node is evicted to disk by looking it up in the cache.  Because the nodes in the cache are organized by a hashtable, it takes $O(1)$ time for the lookup operation. If the node is not in the cache, \toolname queries the attributes of the node and the hopset that contains it. Then \toolname merges them and recalculates the \ac{has}. In this way, \toolname can detect the entire \ac{apt} attack campaigns.

\subsection{Terminal Identification}\label{sec:localfilter}
In Terminal Identification, \toolname scans the in-memory cache and identifies suspicious process nodes as \attcandi based on their node-level features.  Note that although the output of \toolname only contains anomalous processes, it does consider anomalous files and IP addresses. \toolname merges the anomalous files and IP addresses to the processes that access them. The logic behind this design decision is that malicious files and IPs cannot be effective before being accessed by a process. Therefore, focusing on anomalous processes can reduce the duplicated alerts on files and IPs without losing \nodeacc. \toolname analyzes three types of node-level features: the command line that starts a process (command line), the files accessed by a process (files), and IP addresses accessed by a process (network).   
Terminal Identification consists of two steps: First, it embeds process nodes into numerical vectors based on node-level features. Second, it uses a machine learning model to detect anomalies.




\subsubsection{Embedding}\label{subsunsection41}
During Terminal Identification, \toolname first projects node-level features of a process to numerical vectors. On a high level, the embedding of a process is a weighted sum of the embedding vectors of the three node-level features. \toolname chooses to embed node-level features with \ac{nlp} technique to handle unseen patterns in node-level features. The embedding process has two steps. \toolname first embeds command lines, file names, and IP addresses, respectively. Then it combines the embedding of them as the final embedding of a process. 


In the first step, \toolname converts the command lines, file paths, and IP addresses into a sentence of natural language and then uses the document embedding tools in \ac{nlp} to convert the sentence into a vector. The insight behind this design is that command lines contain natural language terms that represent the semantics of a process. For instance, in the command ``date -d 4857 second ago +\%s'', ``date'',  ``second'', and ``ago'', are natural language terms that indicate the functionality of the process. \toolname converts non-alphanumeric symbols into spaces to convert node-level features into sentences. For example, \toolname converts file ``/etc/tmp/log.txt'' into a sentence ``etc tmp log txt'' and converts the quadruple of IP ``$\textless$126.7.8.7, 80, 162.0.0.1, 8080$\textgreater$'' into a sentence ``126 7 8 7 80 162 0 0 1 8080'' by considering ``.'' as spaces. After the conversion, \toolname uses FastText~\cite{bojanowski2017enriching} to convert the sentence to a numerical vector. We choose FastText because of its efficiency~\cite{bojanowski2017enriching}. We can also use other more sophisticated sentence embedding techniques. However, since \toolname has achieved magnitudes higher \nodeacc than baselines in our evaluation, we leave the design of such a better sentence embedding technique as our future work.

The main challenge for embedding node-level features is that they may contain strings that are not a part of natural languages.  For example, ``/var/spool/8b7dc29d0e'' contains the hash string ``8b7dc29d0e''. Existing \ac{nlp}-based document embedding techniques cannot process these special tokens. Therefore, \toolname removes the non-natural-linguistic by deleting tokens that do not have valid meanings with the \ac{nlp} technique Nostril~\cite{Hucka2018}.

The second step for embedding is to sum the numerical vectors of the three node-level features. Formally, the embedding vector of a process is defined as: $$V_{p} = w_{c} * V_{c} + \sum{w_{fi} * V_{fi}} + \sum{w_{ni} * V_{ni}}$$, where $V_{c}$, $V_{fi}$, $V_{ni}$ are the embedding vectors of the command line, files and network connections, respectively, $w_{c}$, $w_{fi}$, and $w_{ni}$ are the weights. 

Formally, the weight $w_{fi}$ of a file is defined as $w = \log(\frac{P}{P_{fi}}) $, where $P$ is the number of all the processes and $P_{f}$ is the number of the processes that operate the file$fi$. We use a similar method to calculate the weights $w_{ni}$ for IP addresses. The logic behind this design is to address files and IP addresses commonly shared by different processes~\cite{liu2018towards}. For example, all processes load the \textit{libc} file. Thus, the \textit{libc} is not useful for modeling the local features of a process. Then, we design the weight to degrade the impact of the files or IP addresses like the \textit{libc} file to improve the accuracy for process modeling. Lastly,   The weight of the command line $w_c$ is the average of the weights of all files and IPs to ensure that  $w_c$ is on the same order of magnitude of files and IP addresses. 


Note that \toolname may achieve better process embedding by leveraging more complex graph embedding techniques~\cite{han2021sigl,2018deep,DBLP:journals/corr/NiepertAK16}. However, since these techniques require passing messages across the whole provenance graph, they are too heavy for an online detection system.

\subsubsection{Anomaly Detection}
\label{sec:localdetect}
To detect \attcandi, \toolname leverages a \ac{vae} model~\cite{kingma2014autoencoding} to calculate the anomaly score for each process node in the provenance graph and then identifies the nodes with higher anomaly scores as \attcandi. \ac{vae} model is widely used as a lightweight anomaly detection model in other tasks~\cite{xu2018unsupervised,lin2020anomaly,ZHOU2021131}. We choose to use the \ac{vae} model because it is efficient for an online detection system~\cite{zhang2021online}. 

 The \ac{vae} model used by \toolname is a standard one for anomaly detection~\cite{standardvae}. The input of the model is $V_{p}$, the embedding vector of a process node $p$. The output is an anomaly score. The high-level process for \toolname is as follows. First, for a process node, \toolname feeds its embedding vector $V_{p}$ to the \ac{vae} model and gets the reconstruction of the input vector as $V'_{p}$. Then, \toolname compares $V'_{p}$ to $V_{p}$. If $V'_{p}$ differs from $V_{p}$ significantly, the input node $p$ is more likely an anomalous node. \toolname numerically measures the difference between $V'_{p}$ and $V_{p}$ with the normalized MSELoss, following other \ac{vae}-based anomaly detectors~\cite{ZHANG2021106748,Papachristodoulou_2021_WACV}. To follow the common terms in machine learning, we use the reconstruction error $RE$ to represent the value of the MSELoss between  $V'_{p}$ and $V_{p}$. 
 Based on existing related work in machine learning~\cite{an2015variational}, $RE$ measures the rareness of a process. In other words, a high $RE$ means the input process node is more likely an anomaly according to historical data. 
 
 
A straightforward method to calculate the anomaly score for a process node is to use the reconstruction error directly. However, this method may generate false positives for unstable processes~\cite{liu2018towards}, which are the processes that often access random files or IP addresses. For instance, a web browser tends to access random IP addresses. Directly using the reconstruction error will constantly identify processes like web browsers as \attcandi, leading to false positives. To address the problem of unstable processes, we introduce a stability score $SV$ to balance the reconstruction error for unstable processes. For \toolname, we define $SV$ as the cluster number of embedding vectors of the processes with the same name as $p$ in the historical data, following related work~\cite{hassan2019nodoze}. 
 Formally, \toolname calculates its anomaly score with the equation: $$AS(p) = \log(\frac{RE(p)}{SV(p)})$$, where $p$ is the embedding vector of a process, $RE$ is the reconstruction error given by the \ac{vae} model, and $SV$ is the stability score of the process $p$. $RE$ is divided by $SV$ to reduce the impact of unstable processes.
 

To detect \attcandi, \toolname marks a process node as anomalous if its $AS$ is higher than the 90th percentile of $AS$ in historical data. This threshold is a widely used threshold for \ac{vae}-based anomaly detection~\cite{s20133738,ZHOU2021131,an2015variational}. In the design of \toolname, we do not use other values for the threshold to ensure that our approach can be generalized. 


\noindent\textbf{Offline Model-Training:}
Although \toolname is an online detection framework, it needs to train the FastText model, the \ac{vae} model, the $SV$ model, and the threshold to raise anomalies from the historical data offline. \toolname trains the FastText model on the command lines, file paths, and IP addresses in the historical data. Then, \toolname uses trained embedding vectors to further train the \ac{vae} model. 
\toolname calculates $SV$ offline by periodically running the DBSCAN algorithm~\cite{khan2014dbscan} on the historical data. It first classifies the process embedding vectors into distinct groups. Then, the process name and its number of clusters are stored in an in-memory hash table for online anomaly score calculation.  

\subsection{Hopset Construction}\label{subsection3}
After detecting \attcandi in Terminal Identification, \toolname runs Hopset Construction to solve the \ac{stp} in the current time window. The hopset in a certain time window is the neighbor context for each terminal, which includes bounded neighbor nodes and the paths to these nodes. \toolname utilizes the greedy algorithm, which we called \ac{isg} algorithm, to construct the hopset based on the local information, such as $AS$ and node degree. Hopset Construction outputs an approximated solution of \ac{stp} with low complexity and bounded competitive ratio, described in Section~\ref{TheoreticalAnalysis}.

Holistically, Hopset Construction follows the algorithm framework in Algorithm~\ref{alg:onlinegreedy} to build the Steiner trees. However, in Algorithm~\ref{alg:onlinegreedy}, finding the shortest paths at line 6 has a complexity of $O(N^2)$ using Dijkstra's algorithm~\cite{hopcroft1983data}, where $N$ is the number of nodes. This is still too expensive for an online detection solution. To this end, we designed an importance-score-guided search that has a complexity of $O(N)$.

The workflow of Hopset Construction is as follows. Assuming that there are $n$ nodes in \attcandi, Hopset Construction first starts $n$ searching procedures, each for one of the \attcandi. The search process is greedy by an importance score of $IV$. Each greedy searching procedure generates a hopset, and the number of nodes is bounded. \toolname stops exploring new nodes if it has already discovered $\theta$ nodes. This early stopping is motivated by the assumption of \insighttwo~\cite{hassan2019nodoze,han2021sigl,wang2020you,277080, zengy2022shadewatcher,9152772, 9152771}, which says that attack actions are topologically close in the provenance graph because attackers need to create the execution chain of attack actions step by step through a sequence of footholds.

Our tool merges overlapping hopsets during greedy searching to efficiently connect terminals within bounded node exploring instead of utilizing the shortest path algorithm with high complexity used in the classic greedy algorithm~\cite{Imase1991DynamicST}. This approximated solution reduces false positives in anomaly detection by identifying topologically close attack-related anomalies. Hopset Construction assigns a \ac{has} to each hopset for future investigation, focusing on more important nodes based on anomaly score and distance from \attcandi. This deprioritizes far nodes to exclude false positives and leverage the attack polymerism of \ac{apt} attacks, which states that attack-related event-level anomalies are topologically close in the provenance graph because attackers conduct attack activities through a few footholds, such as backdoors or reverse shells.

The key component of Hopset Construction is the design of the importance score, which allows \toolname to focus limited computational resources on more important nodes. Our design of $IV$ considers two major factors. The first is the anomaly score $AS$. The second is the distance from the closest node in \attcandi. The intuition behind the distance is to leverage the attack polymerism of \ac{apt} attacks~\cite{han2021sigl,9152772,hassan2019nodoze}, which says that the influence of an attack step candidate decreases with distance. Thus, \toolname deprioritizes the nodes that are far from \attcandi to further exclude false positives in \attcandi. Formally, $IV$ of a node $n$ is defined as: 

\begin{equation}
    IV(n)={\alpha}^i({\beta}*AS(n)+{\gamma}*FANOUT(n))
\label{eq:iv}
\end{equation}

In Equation~\ref{eq:iv}, ${\alpha}^i$ reflects the distance between $n$ and \attcandi. It decreases $IV$ when $n$ is far from \attcandi. Specifically, $0<{\alpha}<1$ is a distance decaying factor, and $i$ is the number of hops between $n$ and its nearest \attcandi. $AS(n)$ is the anomaly score of the node, as defined in Section~\ref{sec:localdetect} 

In Equation~\ref{eq:iv}, we also introduce a supplementary factor $FANOUT(n)$ for a node $n$, which is defined as $FANOUT(n)=out\_degree(n)/(in\_degree(n) + 1)$. We introduce this term because we observe that some processes have the tendency to generate a multitude of ``leaf'' nodes that do not propagate data and control flow dependencies.  These ``leaf'' nodes are not interesting for \ac{apt} attack detection and investigation~\cite{liu2018towards}. Thus, we use the $FANOUT$ term to deprioritize them. We combine  $FANOUT$ and   $AS$ with two weights $\beta$ and $\gamma$. Note that in the design of \toolname, $AS$ is the main factor, while $FANOUT$ is the supplementary factor. Thus, \toolname only requires $\beta >> \gamma$.


\noindent\textbf{Graph Anomaly Score Calculation:} The last step of Hopset Construction is to assign the anomaly score \ac{has} to each hopset. We define the \ac{has} for a hopset as the sum of all anomaly scores of all nodes in the hopset, or formally, $has(H_{i}) = \sum_{n\in H_{i}}{AS(n)}$. Note that since \toolname has already excluded nonimportant nodes based on $IV$, nodes in hopset are likely to be attack-relevant. Thus, the sum of $AS$ only includes the nodes that are highly likely to be attack-relevant. This design helps improve the \nodeacc and \graphacc, defined in Section~\ref{metrics}, by avoiding attack-irrelevant events.

\subsection{Comprehensive Detection}\label{ComprehensiveDetection}

To conduct comprehensive detection, \toolname first updates the cache hopsets in the memory with the hopsets that are constructed in the current time window. \toolname merges the hopsets of current time window with the hopsets in the cache if they have the same nodes. However, if we merge the hopsets directly, in the worst situation, a long-running process was identified as terminal and was updated $\theta$ different neighbors in each time window, which can result in dependency explosion. To prevent it, we limit the hopset of each terminal within $\theta$ nodes and replace the nodes with lower $IV$ values with the ones with higher $IV$ values while merging. 

After hopset merging, \toolname assigns \ac{has} to the updated hopsets, as described in Section~\ref{InmemoryCache}. In this way, we can connect the \ac{stp} solutions of each time window to global solution so that we can reconstruct the long-term attack campaign. Then \toolname leverages the Grubbs's test~\cite{Grubbs-test} to detect hopsets with abnormally high \ac{has}. Grubbs's test detects whether the largest value of a set of samples is an outlier. We choose Grubbs's test because it is non-parameterized and robust to polluted training datasets. \toolname runs the Grubbs's test on the in-memory cache for multiple rounds until no outliers are flagged. Finally, the detected outliers are identified as attack campaigns and alerts will be raised. 

\subsection{Theoretical Analysis}\label{TheoreticalAnalysis}
\noindent\textbf{Complexity:}
In Terminal Identification, \toolname embeds a process to a numerical vector considering the node-level features that the process directly accesses. Thus \toolname needs to explore the node-level features through edges with the complexity $O(E)$.
In Hopset Construction, \toolname constructs hopset for each terminal by greedy exploring $\theta$ nodes. Therefore the complexity of the detection is $O(\theta N)$, where $N$ is the number of nodes in the provenance graph. 

\noindent\textbf{Competitive Ratio:} 
In Section~\ref{subsection3}, we replace the shortest path exploration at line 6 of Algorithm~\ref{alg:onlinegreedy} with the importance-score-guided search. This replacement may result in a larger Steiner tree in the detection, compromising the conciseness. Therefore, in this section, we analyze the new competitive ratio, which is the ratio between the size of the Steiner tree generated by our algorithm to the theoretically optimal solution. Recall that a standard online \ac{stp} optimization algorithm has the competitive ratio of $O(\log(k))$, where $k$ is the number of terminals. Therefore, we only need to compare our approach to the standard online \ac{stp} optimization algorithm.

Formally, in our problem, in time window $i$, \toolname gets a solution of \ac{stp} $S_{i}$, the hopsets in our algorithm. Our object is to minimize the total weight $c(\cup^{k}_{i}S_{i})$, where $k$ is the number of time windows that we have experienced. Assume $S^{*}$ is the standard \ac{stp} solution with cost $c(S^{*}) = STANDARD(G)$ and $O^{*}$ is the standard \ac{stp} solution with cost $c(O^{*}) = OPT(G)$~\cite{RandomizedAlgorithms}. So the competitive ratio can be derived as: $$\frac{c(\cup^{k}_{i}S_{i})}{c(S^{*})}*\frac{c(S^{*})}{c(O^{*})}=\frac{c(\cup^{k}_{i}S_{i})}{c(S^{*})}\log(k)$$

For $\frac{c(\cup^{k}_{i}S_{i})}{c(S^{*})}$, we have the following derivation:
\begin{equation}
\frac{c(\cup^{k}_{i}S_{i})}{c(S^{*})} \le \frac{\sum^{k}_{i}c(S_{i})}{c(S^{*})} \le \frac{|T|\theta}{c(S^{*})} \le \frac{|T|\theta}{|T| - 1} \le 2\theta
\label{eq:competitiveratio}
\end{equation}

As we defined in Section~\ref{sec:overview}, we assign each edge the weight of 1, so the total weight of the solution is the number of edges. For the first step in Equation~\ref{eq:competitiveratio}, the same edges can be merged so we have $c(\cup^{k}_{i}S_{i}) \le \sum^{k}_{i}c(S_{i})$. In each time window, \toolname explores at most $\theta$ nodes for each terminal. Thus, there are at most $\theta$ edges. And we guarantee the hopset of each terminal within $\theta$ nodes so the total number of edges is less than $|T|\theta$, where $|T|$ is the number of terminals. Thus we have $\sum^{k}_{i}c(S_{i}) \le |T|\theta$ for the second step in Equation~\ref{eq:competitiveratio}. And for the weight of optimal solution $c(S^{*})$, it needs at least $|T| - 1$ edges for connection, so we have $c(S^{*}) \ge |T| - 1$ in the third step. Then obviously, $\frac{|T|\theta}{|T| - 1} \le 2$ because $|T| \ge 2$ in most of the cases in \ac{apt} detection. 

To sum up, \toolname can get a $2{\theta}\log(k)=O(\log(k))$ bounded competitive ratio since $\theta$ is a constant. In other words, our importance-score-guided search does not change the worst-case competitive ratio of the original online \ac{stp} algorithm.

\section{Evaluation}\label{section5}
\toolname has three design goals: (1) achieving high \nodeacc, (2) achieving high \graphacc, and (3) achieving high throughput. Therefore we focus on answering the following research questions. 

\begin{itemize}[noitemsep, topsep=1pt, partopsep=1pt, listparindent=\parindent, leftmargin=*]
\item \textbf{RQ 1}: Can \toolname achieve higher \graphacc than SOTA solutions?

\item \textbf{RQ 2}: Can \toolname achieve higher \nodeacc  than SOTA solutions?

\item \textbf{RQ 3}: Can \toolname achieve higher detection efficiency than SOTA solutions?

\item \textbf{RQ 4}: What is the impact of our optimization on efficiency?
\item \textbf{RQ 5}: Can \toolname effectively detect and recover the scenarios of real attacks in a production environment?
\item \textbf{RQ 6}: How do the hyperparameters affect the performance of \toolname?

\end{itemize}
\subsection{Experiment Protocol}
\label{sec:protocol}
We implement \toolname in Python with 5K+ \ac{loc} and integrate it into Sangfor's \ac{edr} solution. \toolname leverages the agents of the industrial \ac{edr} to collect provenance data from monitored hosts. The collected data have the same format as other detection systems~\cite{Han2020,HOLMES}.

We choose HOLMES~\cite{HOLMES} and UNICORN~\cite{Han2020}, two of the representative online provenance-based \ac{apt} attack detection systems, and ProvDetector~\cite{wang2020you}, one of the well-known offline provenance-based for detecting stealthy malware, as the baselines for comparison. HOLMES is one of the \ac{sota} rule-based systems. We use HOLMES as a baseline because it provides complete detection rules in its paper so that we can re-implement it. In addition, there are two other rule-based systems, MORSE~\cite{9152772} and RapSheet~\cite{9152771}, that are as effective as HOLMES. However, both alternatives rely on rules from commercial EDRs or enterprise-specific policies that are not reported in the paper. Thus, we decide not to implement MORSE and RapSheet to avoid possible bias. UNICORN is the \ac{sota} learning-based detection system. We directly use the published source code of UNICORN and use its default parameters, but modify its data parser to accept our data format. In order to more comprehensively evaluate the performance of \toolname, we also choose one of the \ac{sota} offline systems, ProvDetector. We implement ProvDetector and release the code in the same GitHub repository.  \del{\sout{Other learning-based systems are overly expensive that do not support online attack detection~\cite{hassan2019nodoze,liu2018towards}, and are older.}}  Our experiments are carried out on a server running Ubuntu 20.04 64-bit OS with 32-core Intel(R) Xeon(R) CPU E5-2620 v4 @ 2.10GHz, 64GB memory.

The main challenge in evaluating an \ac{apt} detection system is to avoid the problem of ``close-world data'', which means building the solution based on a known dataset that leads to overfitting.  Existing solutions~\cite{han2021sigl,hassan2019nodoze,wang2020you,HOLMES,10.1145/3319535.3363217,9152772,9152771}, including HOLMES~\cite{HOLMES} and UNICORN~\cite{Han2020} are facing this problem. To avoid the ``close-world'' problem, we conduct an open-world experiment by deploying \toolname to realistic production environments, with the parameters obtained from the close-world experiment. The high-level summarization of our datasets is in Table~\ref{tab:datasets}.

\begin{table}[t!]
    \scriptsize
    \centering
    \caption{Summary of our evaluation datasets}
    \label{tab:datasets}

    \begin{threeparttable}
    \resizebox{0.49\textwidth}{!}{  
    \begin{tabular}{|l|l|r|r|r|r|r|} 
    \hline
                                                                          & \textbf{Dataset}     & \textbf{\#\ac{apt}}  & \textbf{Duration}     & \textbf{\#Host}   & \textbf{Event Rate}  & \textbf{\#Activities}\tnote{*} \\ 
    \hline
    \multirow{5}{*}{\begin{tabular}[c]{@{}l@{}}Close\\World\end{tabular}} & DARPA-CADETS         & 3                     & 247h                  & 1                & 16.87 eps   & 21                     \\ 
    \cline{2-7}
                                                                          & DARPA-THEIA   & 1                     & 247h                  & 1          & 11.25 eps        & 97                     \\ 
    \cline{2-7}
                                                                          & DARPA-TRACE       & 2                     & 264h                  & 1         & 75.76 eps         & 93                     \\ 
    \cline{2-7}
                                                                          & Industrial Arena     & 3                     & 336h                  & 22         &  40.74  eps      & 197                    \\ 
    \cline{2-7}
                                                                          & In-lab Arena     & 5                & 144h                  & 5        &  48.23 eps               & 202                    \\ 
    \hline
    \begin{tabular}[c]{@{}l@{}}Open\\World\end{tabular}     & - & 7 & 120h & 300+ & 39.35 eps & 568\\

    \hline
    \end{tabular}}
    \begin{tablenotes}
        \item[*] \#Activities is the number of malicious activities in the dataset.  
    \end{tablenotes}
    \end{threeparttable}
    \end{table}

\subsection{Metrics}\label{metrics}
Formally, we define \graphacc as two metrics: \graphprecision and \graphrecall.

\noindent\textbf{Graph-level accuracy}.
We define \graphprecision as: $\frac{GTP}{GTP+GFP}$, where $GTP$ and $GFP$ are graph-level true positives and false positives, respectively. 
We say a provenance graph is $GTP$ if it contains attack steps and is reported as an alert. 
For instance, the provenance graph in Figure~\ref{fig:motiv} is a $GTP$. Otherwise, we consider it as $GFP$. 
We define \graphrecall as $\frac{GTP}{GTP+GFN}$, where $GFN$ is the number of graph-level false negatives. We say a provenance graph is $GFN$ if it contains attack steps but is not identified as an alert.

\noindent\textbf{Node-level accuracy}.
Node-level accuracy measures the granularity of an online detection system. 
It includes metrics: \nodeprecision and \noderecall.
Specifically, \nodeprecision is defined as $\frac{NTP}{NTP+NFP}$, where $NTP$ and $NFP$ are the numbers of node-level true positives and false positives, respectively. We say a node in a provenance graph is $NTP$ if it is attack-relevant and is included in an alert. Otherwise, we consider it as $NFP$. We define \noderecall as $\frac{NTP}{NTP+NFN}$, where $NFN$ is the number of node-level false negatives. We say an event in a provenance graph is $NFN$ if it is attack-relevant but is not included in an alert.
\subsection{Close-World Experiment}
In the close-world evaluation, we build five datasets. The first three are CADETS, THEIA, and TRACE from the DARPA TC \ac{e3} database~\cite{drapa}. We choose these three datasets because they are widely used in evaluating provenance-based detection systems~\cite{Han2020,9152772}. Besides the three datasets from DARPA TC, we also build two datasets, industrial Arena and In-lab Arena, that represent the realistic production environment of a security company, \anosec. The Industrial Arena dataset includes the provenance data of a 14-day internal security evaluation engagement of \anosec. It contains 22 working machines of its employees and three \ac{apt} attacks simulated by a professional red team. The three attacks include the well-known APT29\cite{apt29} attack, the APT32\cite{apt32} attack, and a new one that exploits the latest Log4j2 (CVE-2021-44228)\cite{Log4jCVE-2021-44228} vulnerability.

In addition to the Industrial Arena dataset, we also built a smaller testbed that simulated the internal environment of \anosec and made the collected data publicly available\footnote{\url{https://github.com/Nodlink/Simulated-Data}}. The data were collected from five hosts: one Ubuntu 20.04 server, two Windows Server 2012 R2 Datacenter, one Windows Server 2019 Datacenter, and one Windows 10 desktop host. We deployed Apache and PostgreSQL on Windows Servers and Nginx and PostgreSQL on Ubuntu 20.04 to simulate servers in \anosec. The Windows 10 desktop was used to simulate the PCs used by the employees. On Windows Server 2012 and Ubuntu 20.04, we carried out the two real attacks from \anosec. The attacker exploits the Apache Struts2-046 (CVE-2017-5638) vulnerability~\cite{struts2-046} for remote code execution and then penetrates the database on Ubuntu 20.04 by exploiting weak passwords. On Windows Server 2012, the attacker exploits the “phpstudy” backdoor and uses “PAExec.exe” to control other Windows servers remotely. On Windows 10, we carried out APT29~\cite{apt29}, FIN6~\cite{FIN6} and SideWinder~\cite{SideWinder} using the Atomic Red Team~\cite{secbudget} and techniques based on MITRE's ATT\&CK~\cite{mitre}. We have also provided detailed attack steps in the documents of our repository.

For each dataset, we partition its benign data into a training set (80\% of the graphs) and a test dataset (20\% of the graphs). We then trained the detection model on the benign training set and evaluated the performance of \toolname on the test set and the attack data.

To build the ground truth for the three datasets from DARPA TC, we first labeled the attack according to the documents provided by DARPA.
We leverage the red team in \anosec to label the attack steps they have made for the Industrial Arena Dataset and the In-lab Arena Dataset. 
\subsection{Open-World Experiment}

In open-world evaluation, we integrate \toolname and baselines to a beta version of \ac{edr} of \anosec and deploy them to monitor the systems of realistic customers. Before our experiment, we have no prior knowledge about the data of customers. Thus, we can avoid over-fitting the test data. We fine-tune the hyperparameters based on the DARPA dataset in ``close-world data'' for \toolname and baselines, and deploy the fine-tuned models to the ``open-world data'' directly. The open-world evaluation evaluates how well can \toolname detect real attacks in production environments.

Our open-world evaluation includes 300+ servers and working machines of employees of 10 industrial customers from \anosec, including schools, research institutes, factories, and healthcare providers. The servers include both Linux and Windows servers that host databases and back-end web services, while the working machines are Windows desktops used by employees in their daily work. Overall, there are 50+ Linux machines and 250+ Windows machines. We monitor the customers for five days. We use the first three days to train the detection model for \toolname and UNICORN,  and use the last two days for testing. 

\noindent\textbf{Ground Truth Building:} 
It is particularly challenging to build the ground truth for the open-world evaluation. Due to the large number of monitored hosts and the open environment, we cannot manually examine all system events to identify which event is attack-relevant. To this end, we first leverage the conventional \ac{edr} with 1000+ rules customized for the customers of \anosec to label the possible attack-relevant events. However, this conventional \ac{edr} is overly conservative. It reports about 135,700 alerts for the two-day testing period. A professional security team of \anosec manually examined the alerts and identified about 2,000 true alerts, which we consider attack-relevant events.  Note that, unlike the \toolname and the two baselines, conventional \ac{edr} cannot automatically reconstruct the attack campaigns from attack-relevant events. Therefore, the same security team manually reconstructs seven \ac{apt} attack campaigns from the attack-relevant events. We put a detailed description of the attacks in Section~\ref{sec:casestudy}.

\noindent\textbf{Hyper-parameter Setting:}
We empirically set the five hyperparameters, $\alpha$, $\beta$, and $\gamma$, $\theta$, and $\epsilon$, of \toolname with DARPA datasets (CADETS, THEIA, and TRACE). We set $\alpha$ as 0.9, ($\beta$,$\gamma$) as (100,1), $\theta=10$, and $\epsilon=0.8$. We will discuss our rationale for choosing these values in Section~\ref{sec:parameter}.

\begin{table}[t!]
    \scriptsize
    \centering
    \caption{Ground truth and detection results of the datasets. The precision and recall of UNICORN for CADETS and THEIA are from the original paper~\cite{Han2020}. ProvDetector fails to report any alerts on the DARPA-CADETS dataset and cannot conduct detection on DARPA-TRACE dataset due to memory limitation. HOLMES fails to report any alerts on the open-world dataset.}
    \resizebox{0.49\textwidth}{!}{\begin{tblr}{      
        cell{1-2}{1-9} = {c},
        cell{3-8}{1} = {l},
        cell{3-8}{2-9} = {r},
        vline{1,2,4,6,8,10},
        stretch=0}
    \hline
    \SetCell[r = 2]{c} \textbf{Dataset}  &
    \SetCell[c = 2]{c} \textbf{ProvDetector} & & \SetCell[c = 2]{c} \textbf{HOLMES} & & \SetCell[c=2]{c} \textbf{UNICORN} & & \SetCell[c = 2]{c} \textbf{\toolname}  \\  \cline{2-9}
& \textbf{P} & \textbf{R}& \textbf{P} & \textbf{R} & \textbf{P} & \textbf{R} & \textbf{P} & \textbf{R}\\ \hline
    DARPA-CADETS              &  NA  &  NA  &  0.03 & 1.00 & 1.00 &  1.00 & 1.00 & 1.00                 \\ 
    \hline
    DARPA-THEIA               &  0.05  & 1.00  & 1.00 & 1.00 & 1.00 &   1.00 & 1.00 & 1.00                           \\ 
    \hline
    DARPA-TRACE                & NA &  NA    & 0.15 & 1.00 & 0.28&  1.00 &  0.67& 1.00                            \\ 
    \hline
    Industrial Arena      &  0.75   &  1.00      &0.04  & 1.00 & 0.67 & 1.00 &  1.00 & 1.00                         \\ \hline
    In-lab Arena           &  0.65 &   1.00     &0.04 & 1.00 & 0.50 & 1.00 & 1.00 &1.00                         \\ \hline
    Open-World             & 0.11 & 1.00  & NA & NA & 0.15 & 1.00 &  0.35 & 1.00                      \\ \hline
    \end{tblr} }
    \label{tab:graphacc}
\end{table}
\subsection{RQ 1: Graph-Level Accuracy}\label{rq1}

We report the \graphprecision and \graphrecall of \toolname and the three baselines in Table~\ref{tab:graphacc}.

\noindent\textbf{Close-World Result:} \del{\sout{As shown in Table~\ref{tab:graphacc}, although both UNICORN and HOLMES can detect all attacks, they report more false positives than \toolname. Specifically, UNICORN and HOLMES generate 14 and 416 $GFPs$, respectively. On the contrary, \toolname only reports one false positive in these datasets. }}
As shown in Table~\ref{tab:graphacc}, although these systems can detect all attacks, they report more false positives than \toolname. Specifically, ProvDetector, UNICORN and HOLMES generate 783, 14 and 416 $GFPs$. On the contrary, \toolname only reports one false positive in these datasets.

ProvDetector fails to detect the attack in DARPA-CADETS and cannot conduct detection on DARPA-TRACE due to memory limitation. It also has the lowest  \graphprecision on DARPA-THEIA because it cannot handle browser activities that explode dependencies. For example, ProvDetector mistakenly reports false alerts when Firefox connects to new IP addresses. UNICORN has lower \graphprecision because it over-approximates the provenance graph. UNICORN projects a provenance graph to a numerical vector for detection. This step downgrades its \graphprecision. HOLMES has more $GFPs$ because its rules are overly conservative. For example, the command-line utility rule of HOLMES marks a process as anomalous if the process runs command-line utilities after accessing an untrusted IP address. This rule generates a lot of false positives for long-running processes such as Nginx: once an Nginx process receives a connection from the untrusted IP, all shell command executions forked by the Nginx process are marked as anomalies. Although building more complex rules can improve the precision of HOLMES, it is tedious to consider the heterogeneity of systems and the number of different types of system activities.

\noindent\textbf{Open-World Results:} We also show the result of the open-world experiment in Table~\ref{tab:graphacc}. Overall, the result is consistent with that of the close-world experiment. \toolname has a lower \nodeprecision in the open-world experiment because the signal-noise ratio is much lower in the open-world experiment. \toolname outperforms the baselines. Particularly, HOLMES FAILS to detect any attacks in the open-world experiment because its rule set lacks rules to detect webshells and process hijacking attacks. Unfortunately, the attacks in the open-world experiment all leverage webshells and related attacks that hijack running processes to stay stealthy. Thus, HOLMES fails to capture any of them.

\subsection{RQ 2: Node-Level Accuracy}\label{rq2}

We evaluate the \nodeacc of \toolname and compare it to our two baselines. We show the results of \nodeprecision and \noderecall in Table~\ref{tab:attack-node-precision} and Table~\ref{tab:attack-node-recall}.

\begin{table}[!t]
    \small
        \centering
        \caption{Node-level precision of \toolname and baselines. PI, HI and UI mean the improvement of \toolname over ProvDetector, HOLMES and UNICORN respectively.}
        \resizebox{0.49\textwidth}{!}{\begin{tabular}{l|rrrc} 
            \hline
                                  & \multicolumn{4}{c}{\textbf{Node-level Precision}}                                                                                             \\ 
            \cline{2-5}
            \multicolumn{1}{l|}{} & \textbf{ProvDetector} & \textbf{HOLMES}     & \textbf{UNICORN}    & \begin{tabular}[c]{@{}c@{}}\textbf{}\\\textbf{NodLink}\\\textbf{(PI,HI,UI)}\\\textbf{}\end{tabular}  \\ 
            \hline
            DARPA-CADETS      &   NA   & $2.84 \times 10^{-3}$ & $1.25 \times 10^{-4}$ & \begin{tabular}[c]{@{}c@{}}0.14\\(-,47,1082)\end{tabular}                                               \\ 
            \hline
            DARPA-THEIA       &  0.01  & $3.61 \times 10^{-3}$ & $1.86 \times 10^{-4}$ & \begin{tabular}[c]{@{}c@{}}0.23\\(23,62,1218)\end{tabular}                                               \\ 
            \hline
            DARPA-TRACE        &  NA & $1.35 \times 10^{-3}$ & $3.20 \times 10^{-5}$ & \begin{tabular}[c]{@{}c@{}}0.25\\(-,184,7817)\end{tabular}                                               \\ 
            \hline
            Industrial Arena    & 0.14 & $5.10 \times 10^{-3}$ & $1.39 \times 10^{-3}$ & \begin{tabular}[c]{@{}c@{}}0.21\\(2,41,152)\end{tabular}                                               \\ 
            \hline
            In-lab Arena        & 0.16 & $8.76 \times 10^{-3}$ & $1.95 \times 10^{-3}$ & \begin{tabular}[c]{@{}c@{}}0.17\\(1,19,87)\end{tabular}                                               \\
            \hline            
            Open-World       & 0.13 & NA & $3.61 \times 10^{-4}$ & \begin{tabular}[c]{@{}c@{}}0.14\\(1,NA,390)\end{tabular}                                               \\
            \hline
            \end{tabular}}
        \label{tab:attack-node-precision}
    \end{table}
    
\begin{table}[!t]
\small
    \centering
    \caption{Node-level recall of \toolname and baselines. Higher is better. A value of 1 means that all attack actions have been captured, while 0 means all attack steps are missed.}
    \resizebox{0.47\textwidth}{!}{\begin{tblr}{
        cell{1-2}{1-4} = {c},
        cell{3-8}{1} = {l},
        cell{3-8}{2-4} = {r},
        vline{2},
        stretch=0}
    \hline
    \SetCell[r=2]{c}& \SetCell[c=3]{c}\textbf{Node-level Recall}& &    \\ \hline
    & \textbf{ProvDetector} & \textbf{HOLMES}  &\textbf{\toolname}  \\ \hline
    DARPA-CADETS &NA & 0.95 & 1.00  \\ \hline
    DARPA-THEIA & 1.00 &0.98  & 1.00\\ \hline
    DARPA-TRACE & NA &0.74  &0.98 \\ \hline
    Industrial Arena & 0.20 & 0.23  & 0.96 \\ \hline
    In-lab Arena & 0.98 & 0.32 & 0.92 \\ \hline
    Open-World & 1.00 & NA  & 1.00 \\ \hline
    \end{tblr}}
    \label{tab:attack-node-recall}
\end{table}

\noindent\textbf{Close-World Results:} The \nodeprecision of \toolname is comparable with offline solution, ProvDetector. For online solutions,  \toolname is one to two orders of magnitude higher than HOLMES and two to three orders of magnitude higher than UNICORN.

For \noderecall, \toolname captures most of the attack steps in its reported provenance graph. On average, \toolname covers 98\% of the attack-relevant events.  The missed steps are about reconnaissance, such as running ``whoami'', ``ipconfig'', ``tasklist'' and ``systeminfo'' to collect information about a system. \toolname has captured all steps that are related to running attack payloads and lateral movement in our experiment. 
HOLMES has a lower node-level recall because it lacks rules to detect several attack steps. For example, it cannot detect attack steps that are initiated by internal files.

We cannot measure the node-level recall for UNICORN because it simply reports all events in a provenance graph regardless of whether an event is relevant or irrelevant to an attack. Thus, although UNICORN captures all attack steps in its reported provenance graphs, it has magnitudes lower \nodeprecision than others. In addition, UNICORN is incompatible with other investigation techniques because all investigation techniques need suspicious nodes or edges of \ac{ioc}s to initiate the investigation. UNICORN does not produce an \ac{ioc} as a starting point for investigation techniques. Instead, it generates a provenance graph that includes attacks. \del{\sout{It will be difficult for security admins to investigate the alerts generated by UNICORN due to their large sizes. }}Therefore, we cannot combine UNICORN with post-processing steps to increase its \nodeprecision.

\noindent\textbf{Open-World Results:} We also report the results of the open-world evaluation in Table~\ref{tab:attack-node-precision} and Table~\ref{tab:attack-node-recall}. We skip the data of HOLMES as it fails to detect any attacks in the open-world experiment. Generally, the results are consistent with the results of close-world experiments. \toolname achieves the same \noderecall as UNICORN with two orders of magnitude higher \nodeprecision.

\subsection{RQ 3: Efficiency}
We evaluate how efficiently \toolname can detect \ac{apt} attacks in a timely manner by measuring its throughput, which is defined as how many system events can be processed per second. Before running the empirical experiments, we first report our theoretical analysis of the complexity of our baselines. 
ProvDetector first extracts all the paths in the provenance graph and calculates the regularity score for each event. Then it converts the detection to finding $K$ longest paths on a directed acyclic graph, which is solved by Epstein's algorithm~\cite{365697} with time complexity $O(E + N\log N)$.
HOLMES processes provenance graphs through a policy matching engine. Therefore, its time complexity is $O(E)$, where $E$ is the number of edges. UNICORN constructs graph histograms with the complexity $O(R*E)$ and uses HistoSketch~\cite{8215527} to generate graph sketches in real time, where $R$ is the number of hops, and the $E$ is the number of edges, as it is claimed in its paper~\cite{Han2020}. Remember that the complexity of \toolname is $O({\theta}N)$. Thus, our analysis expects that \toolname and HOLMES are much faster than ProvDetector and UNICORN.

\begin{figure}[t!]
    \centering
    \includegraphics[width=0.48\textwidth]{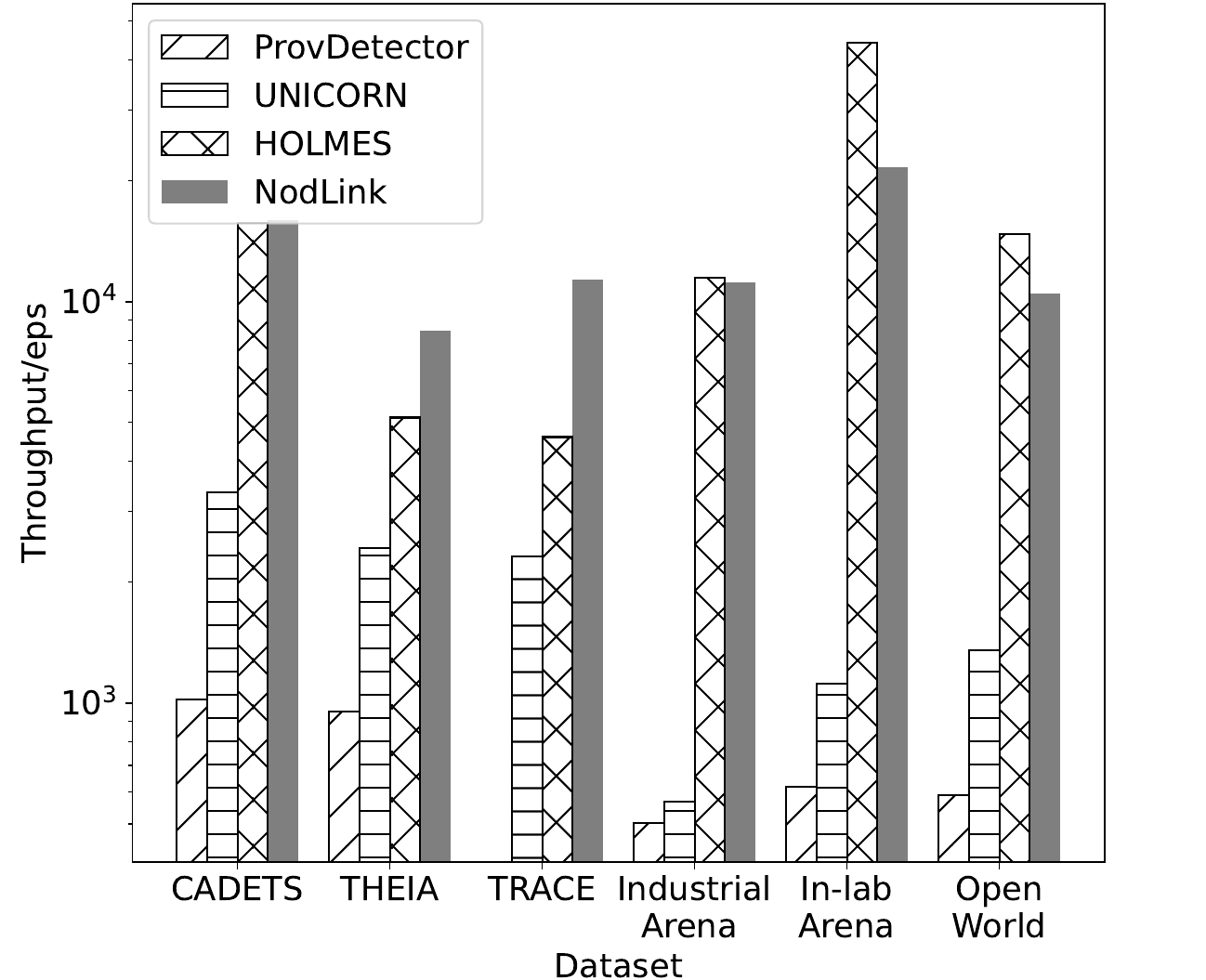}
    \caption{Throughput of \toolname and the three baselines on different datasets. ProvDetector failed to run on DARPA-TRACE, so we cannot get the throughput.}
    \label{fig:throughput}
\end{figure}

Figure~\ref{fig:throughput} shows the empirical results. The y-axis is the number of events processed per second in each dataset. Overall, the throughput of \toolname is comparable to the throughput of HOLMES. \toolname has a higher throughput in DARPA-THEIA and DARPA-TRACE, while HOLMES has a higher throughput for the In-lab Arena dataset and the open-world experiment. 
HOLMES operates in two stages: (1) match \ac{ttp} from provenance data, (2) link \ac{ttp} to \ac{hsg}. However, HOLMES fails to match some crucial steps in the in-lab arena and open-world datasets, as illustrated in Table~\ref{tab:attack-node-recall}. This failure at the first stage eliminates the need for the second step, resulting in a higher throughput performance of HOLMES in these two datasets.
ProvDetector shows the poorest throughput because it needs to calculate the regularity score for each event and find $K$ longest paths.
As listed in Table~\ref{tab:datasets},  an open-world host generates an average of 40 events per second. In industrial settings, a central-based detection system is expected to handle hundreds to thousands of hosts in a cluster~\cite{dong2023yet}. With an assumption of 500 hosts, the system must process 20,000 events per second. Therefore, the throughput of UNICORN and ProvDetector falls short of industrial requirements.
Overall, our evaluation shows that \toolname has a comparable or higher throughput than existing detection systems.

\subsection{RQ 4: Ablation Study on Efficiency}

\toolname designs in-memory cache and the \ac{isg} to improve efficiency. In this section, we conduct an ablation study on their impact respectively on efficiency using DARPA datasets. To evaluate the cache design, we disable the cache and store all the provenance data and cached graphs in the graph database Neo4j. To evaluate the \ac{isg} algorithm, we use the conventional Steiner Tree approximation algorithms, Kou~\cite{kou1981fast}, with $O(|T||V|^{2})$ time complexity, and Mehlhorn~\cite{ZELIKOVSKY199379} with $O(|E| + |V|\log(|V|))$ time complexity, in each time window as the baseline. We show the results in Table~\ref{tab:ablation}. We set the throughput of \toolname as the baseline and calculate the ratio between them. Overall, the design of the in-memory cache and the \ac{isg} search algorithm improves the detection efficiency greatly.

As shown in Table~\ref{tab:ablation}, when disabling the in-memory cache design, the efficiency is 82.54 - 159.06 times slower than the original design. When replacing our algorithm with Kou, the efficiency is 8406.86 times slower on the DARPA-CADETS. We failed to get the exact time of Kou due to the long processing time for the other two datasets. When replacing our algorithm with Mehlhorn, the efficiency is 3.08 - 111.21 times slower. We can find that the efficiency of these algorithms decreases as the dataset size increases, which is a consequence of their high complexity.

\begin{table}[t!]
    \small
        \centering
        \caption{Ablation study on the efficiency of \toolname. ``w/o Cache'' disables the in-memory cache. ``w/Kou'' replaces \ac{isg} with Kou algorithm. ``w/Mehlhorn'' replaces \ac{isg} with Mehlhorn algorithm. We regard the throughput of \toolname as standard. }
        \resizebox{0.49\textwidth}{!}{\begin{tabular}{l |r  r r  r }
        \hline
         & \begin{tabular}[c]{@{}c@{}}\toolname\\ w/o Cache \end{tabular} & \begin{tabular}[c]{@{}c@{}}\toolname\\ w/Kou \end{tabular}   &\begin{tabular}[c]{@{}c@{}}\toolname\\w/Mehlhorn \end{tabular}   & \begin{tabular}[c]{@{}c@{}}\toolname\\w/\ac{isg} \end{tabular}  \\\hline
        DARPA-CADETS & $\times$82.54 & $\times$8406.86 & $\times$3.08 & 1 \\ \hline
        DARPA-THEIA & $\times$102.36 & - & $\times$47.79 & 1\\ \hline
        DARPA-TRACE & $\times$159.06& - & $\times$111.21 & 1 \\ \hline
        \end{tabular}}
        \label{tab:ablation}
    \end{table}

\subsection{RQ 5: Attacks Detected In Production}\label{sec:casestudy}
In this section, we show how \toolname detects real attacks in the open-world experiment with seven case studies. Overall, HOLMES fails to detect all these attacks because it lacks the rule to detect the code injection of webshell. \toolname and UNICORN can successfully detect them. However, \toolname has a much higher \nodeprecision, which can always be magnitudes larger than UNICORN.

\noindent \textbf{Attack 1: }In this attack, the attacker first hijacks the Windows IIS Web Server ``w3wp.exe'' through web shell injection. Then the attacker uses ``csc.exe'' to execute a trojan. In addition, the attacker also uploads the remote tools ``GotoHTTP\_x64.exe'' to edit the registry and manages the devices for persistence and privilege escalation. Finally, he leaves a backdoor ``Wlw.exe'' to carry out intranet blasting with ``fscan.exe'' and uses ``wevtutil'' to clear footprints.

\begin{figure}[t!]
    \centering
    \includegraphics[width=0.5\textwidth]{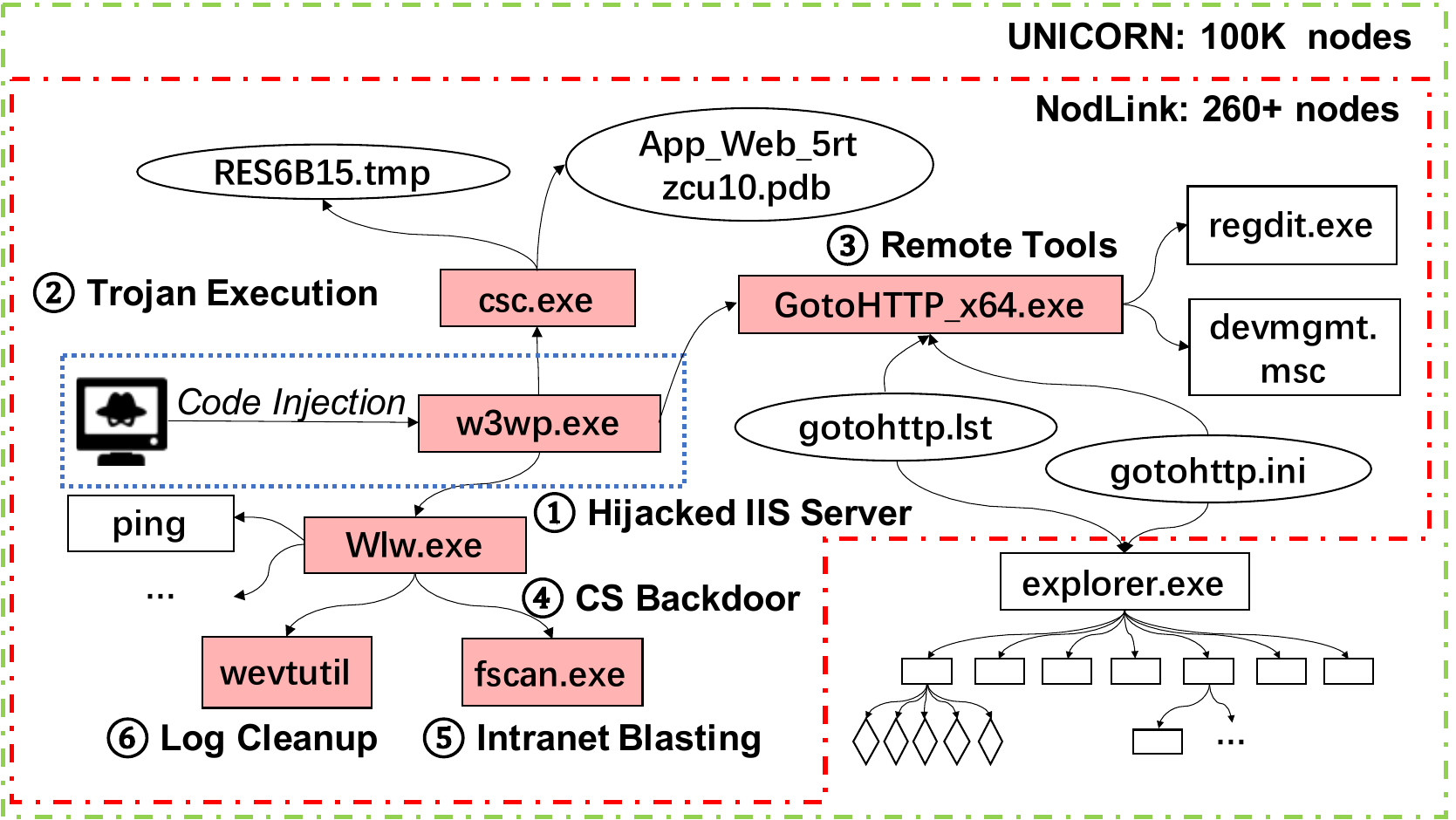}
    \caption{An attack that first injects a web shell to the Windows IIS Web server and leaves a \ac{cs} backdoor for lateral movement. Then it uploads the remote invocation tool for persistence and privilege escalation.}
    \label{fig:C3}
\end{figure}

\toolname and UNICORN have successfully detected this attack. However, \toolname has a much higher \nodeacc. Figure~\ref{fig:C3} shows the provenance graph for the attack.  The part boxed by the red dashed lines is the provenance graph generated by \toolname (with about 260 nodes) and the part boxed by green dashed lines is generated by UNICORN (with more than 100K nodes). The \nodeprecision of \toolname in this case is 0.27, which is magnitudes larger than UNICORN. Besides, \toolname can also identify the core attack steps, such as running the webshell (``w3wp.exe''), the trojan ``csc.exe'', and other attack tools, as \attcandi, which are marked as red-solid boxes. In Figure~\ref{fig:C3}, we can observe that the provenance graph of UNICORN has many attack-irrelevant events caused by the file explorer of Windows (``explorer.exe''). These events are problematic for attack investigation.  HOLMES fails to detect this attack because it lacks the rule to detect the code injection of webshell.

Our two baselines HOLMES and UNICORN cannot perform well on this attack scenario. For HOLMES, the code injection in the initial compromise stage, labeled with blue dotted boxes in Figure~\ref{fig:C3}, cannot be matched up with the rules listed in HOLMES. Therefore it misses the whole attack. For UNICORN, although it can detect the attack, it reports the whole provenance graph with more than 100,000 nodes, which contains a large number of irrelevant events, such as the file manager application ``explorer.exe''. Thus, it is very difficult for security admins to analyze the result of UNICORN.

\begin{figure}[!t]
    \centering
    \includegraphics[width=0.5\textwidth]{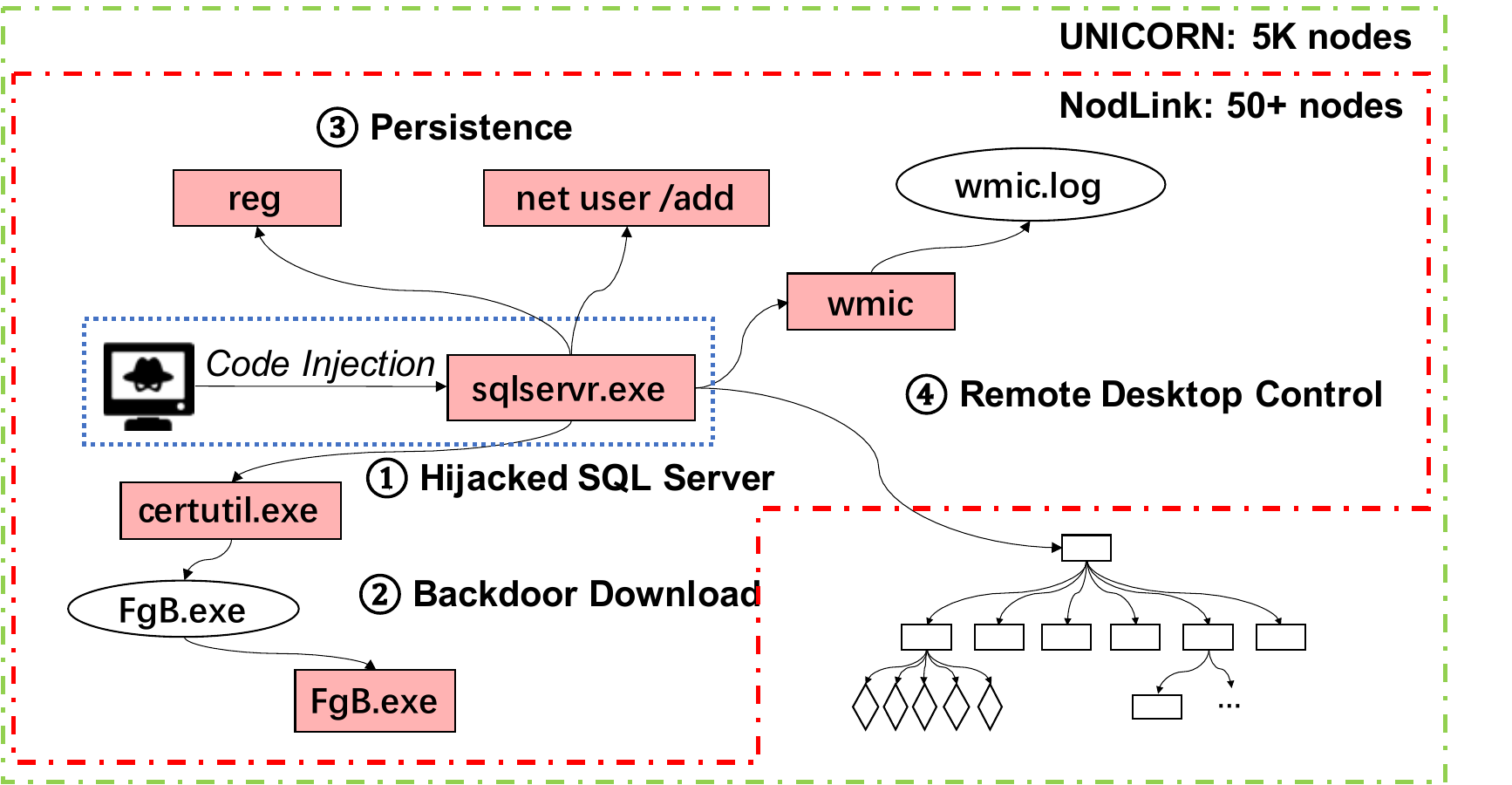}
    \caption{An attack that hijacks the SQL server and downloads a backdoor for lateral movement. Then it adds the user to administrators and sets up the remote desktop control.}
    \label{fig:C1}
\end{figure}

\noindent \textbf{Attack 2: }As shown in Figure~\ref{fig:C1}, the attacker first hijacks the SQL Server on one host from an internal IP, then uses ``certutil'' to download a backdoor ``FgB.exe'', and leverages ``wmic'' to set up the remote desktop control. Finally, the attacker adds a hidden user to the administrators for persistence, accesses the data stored in the database, and sends them to another internal IP for data exfiltration. 

The part boxed by the red dashed lines is the provenance graph generated by \toolname (with about 50 nodes) and the part boxed by green dashed lines is generated by UNICORN (with more than 5K nodes). The \nodeprecision of \toolname in this case is 0.34. Besides, \toolname can also identify the core attack steps, such as hijacked SQL Server (``sqlservr.exe''), the backdoor ``FgB.exe'', and other \ac{lol} attack tools, as \attcandi, which are marked as red-solid boxes. In Figure~\ref{fig:C1}, we can observe that the provenance graph of UNICORN has many attack-irrelevant events that are generated by the normal behaviors of the SQL Server process (``sqlservr.exe''). 

\begin{figure}[t!]
    \centering
    \includegraphics[width=0.5\textwidth]{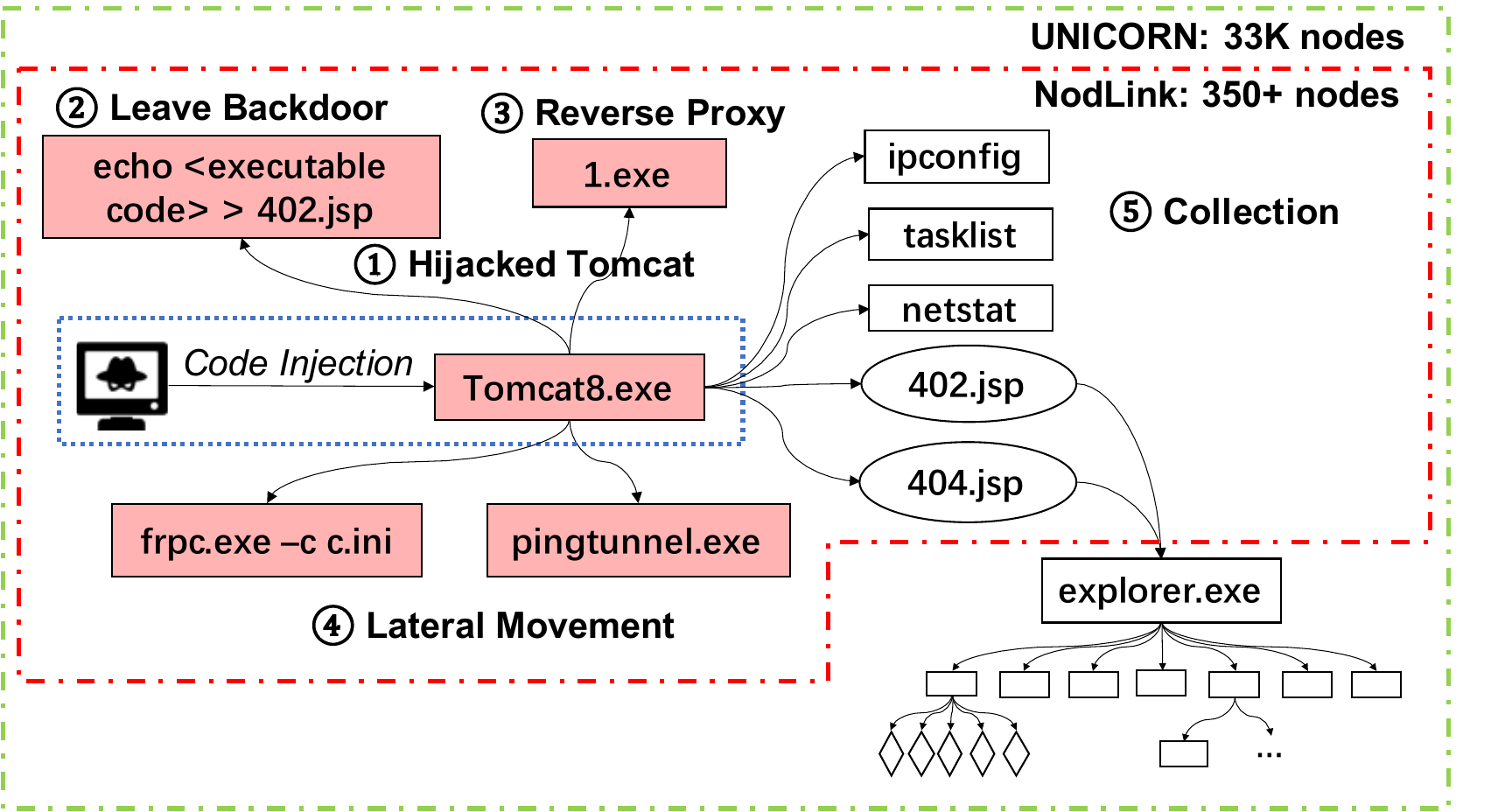}
    \caption{An attack that injects executable code to Tomcat and leaves a backdoor. Then it uploads the reverse proxy tool, carries out lateral movement and collects system information along with the whole procedure.}
    \label{fig:C2}
\end{figure}

\noindent \textbf{Attack 3: }As shown in Figure~\ref{fig:C2}, in this attack, the adversary injects a web shell to a Tomcat server ``Tomcat8.exe'' and leaves several backdoors. Then he uses ``frpc.exe'' and ``pingtunnel'' for lateral movement, runs a reverse proxy ``1.exe'' to keep C\&C and scans for sensitive information.

The part boxed by the red dashed lines is the provenance graph generated by \toolname (with about 350 nodes) and the part boxed by green dashed lines is generated by UNICORN (with more than 33K nodes). The \nodeprecision of \toolname in this case is 0.22. Besides, \toolname can also identify the core attack steps, such as hijacked Tomcat (``Tomcat8.exe''), process that leaves a backdoor, the reverse proxy ``1.exe'', and other steps for lateral movement, as \attcandi, which are marked as red-solid boxes. In Figure~\ref{fig:C2}, we can observe that the provenance graph of UNICORN has many attack-irrelevant events that are generated by the file explorer of Windows (``explorer.exe''). 

\begin{figure}[t!]
    \centering
    \includegraphics[width=0.52\textwidth]{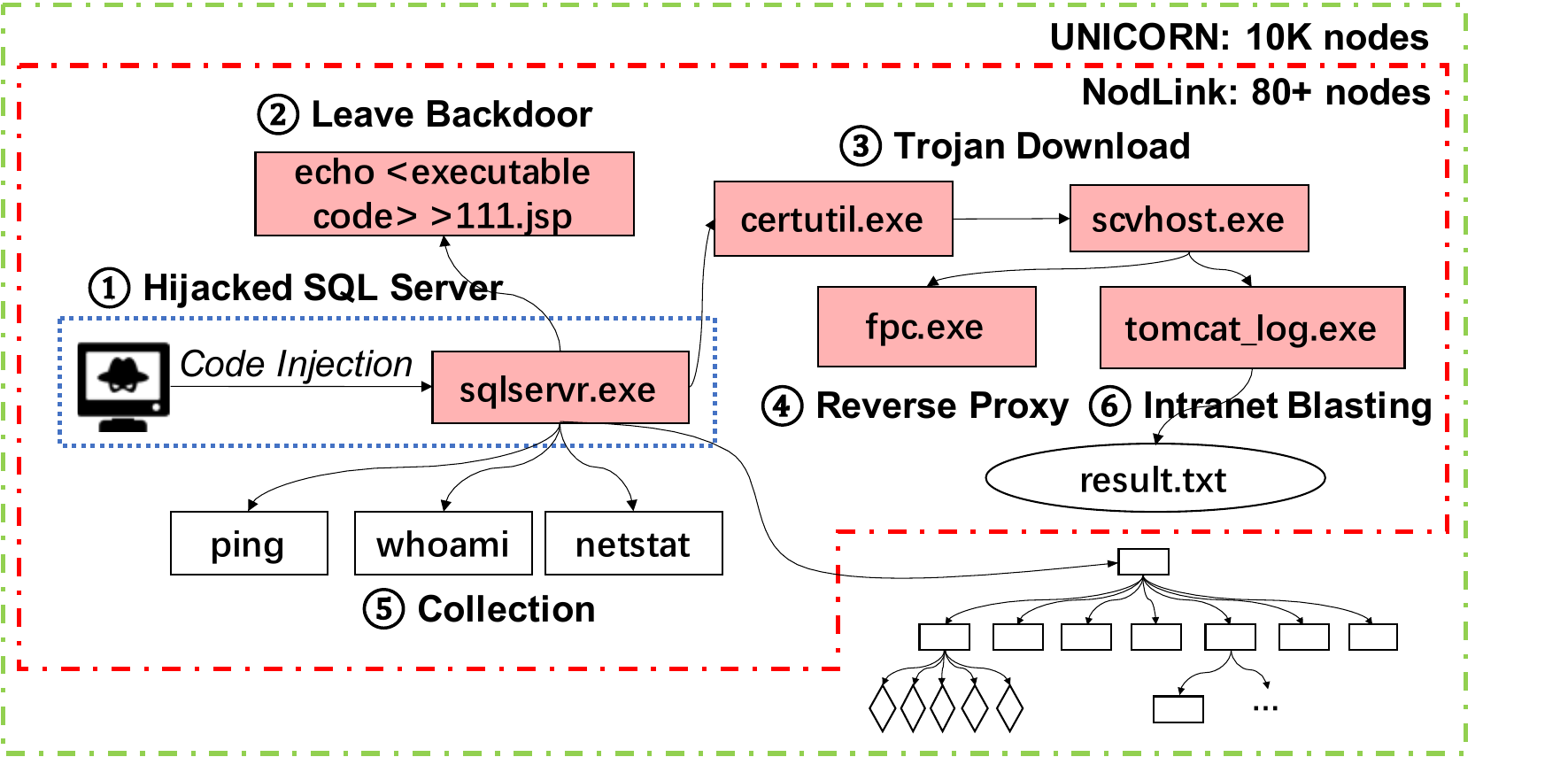}
    \caption{An attack that first injects a web shell to SQL Server and leaves a backdoor. Then it downloads Trojan for reverse proxy and intranet blasting. During the whole procedure, it collects the system information.}
    \label{fig:C4}
\end{figure}

\noindent \textbf{Attack 4: }As shown in Figure~\ref{fig:C4}, the attacker hijacks the SQL Server with backdoor ``111.jsp'' and uses ``certutil.exe'' to download ``scvhost.exe'', the \ac{cs} Trojan with a deceptive process name that is similar to the Host Process for Windows Services ``svchost.exe'', on the server, and successfully communicates to the outside. Then he sets up a reverse proxy, collects the sensitive information and uses ``fscan.exe'' which is renamed as ``tomcat\_log.exe'' to scan the intranet. It generates a ``result.txt'' file to record multiple intranet vulnerabilities.

The red dashed box contains the provenance graph from \toolname (with about 80 nodes), while the green dashed box holds the UNICORN-generated part (over 10K nodes). The \nodeprecision of \toolname in this case is 0.75. \toolname can also identify the core attack steps, such as hijacked SQL Server (``sqlservr.exe''), process that leaves a backdoor, the Trojan process ``scvhost.exe'', and other steps for lateral movement and reverse proxy, as \attcandi, which are marked as red-solid boxes. In Figure~\ref{fig:C4}, we can observe that the provenance graph of UNICORN has many attack-irrelevant events that are generated by the normal behaviors of the SQL Server process (``sqlservr.exe''). 

\begin{figure}[t!]
    \centering
    \includegraphics[width=0.52\textwidth]{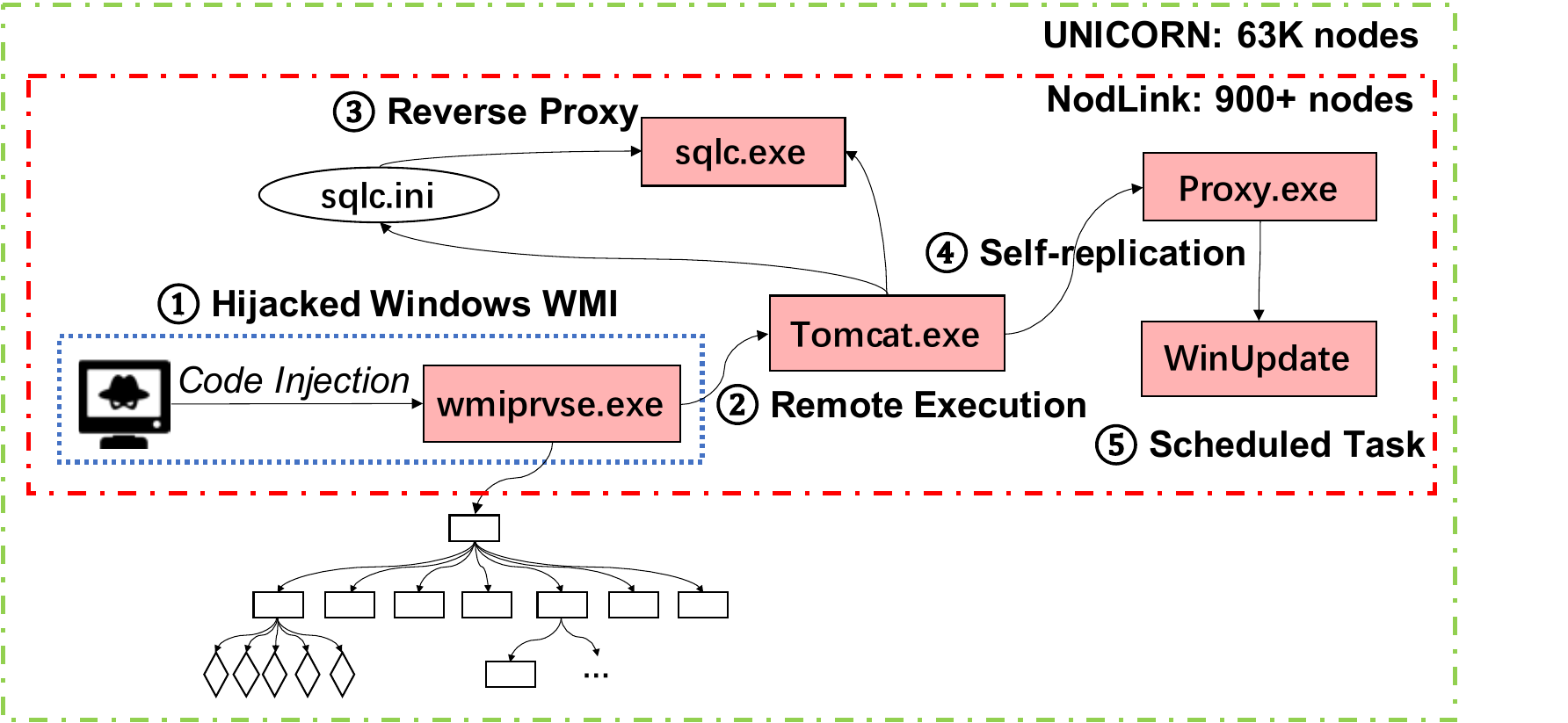}
    \caption{An attack that hijacks the \ac{wmi} process and executes Tomcat remotely. Then it sets up reverse proxy and copies itself to create scheduled task for permission maintenance. }
    \label{fig:C5}
\end{figure}

\noindent \textbf{Attack 5: }
As shown in Figure~\ref{fig:C5}, the attacker uses ``wmiprvse.exe'', \ac{wmi} that provides management information and control in an enterprise environment, to run remote commands on the host through the domain account and execute the ``Tomcat.exe'' to copy itself to ``Proxy.exe'' and sets up reverse proxy tools ``sqlc.exe''. Then the adversary creates scheduled tasks ``WinUpdate'' for permission maintenance.

The part boxed by the red dashed lines is the provenance graph generated by \toolname (with about 900 nodes) and the part boxed by green dashed lines is generated by UNICORN (with more than 63K nodes). The \nodeprecision of \toolname in this case is 0.22. Besides, \toolname can also identify the core attack steps, such as hijacked \ac{wmi} (``wmiprvse.exe''), the remote execution ``Tomcat.exe'', and other steps for reverse proxy and scheduled task, as \attcandi, which are marked as red-solid boxes. In Figure~\ref{fig:C5}, we can observe that the provenance graph of UNICORN has many attack-irrelevant events that are generated by the normal behaviors of the SQL Server process (``sqlservr.exe''). 

\begin{figure}[t!]
    \centering
    \includegraphics[width=0.5\textwidth]{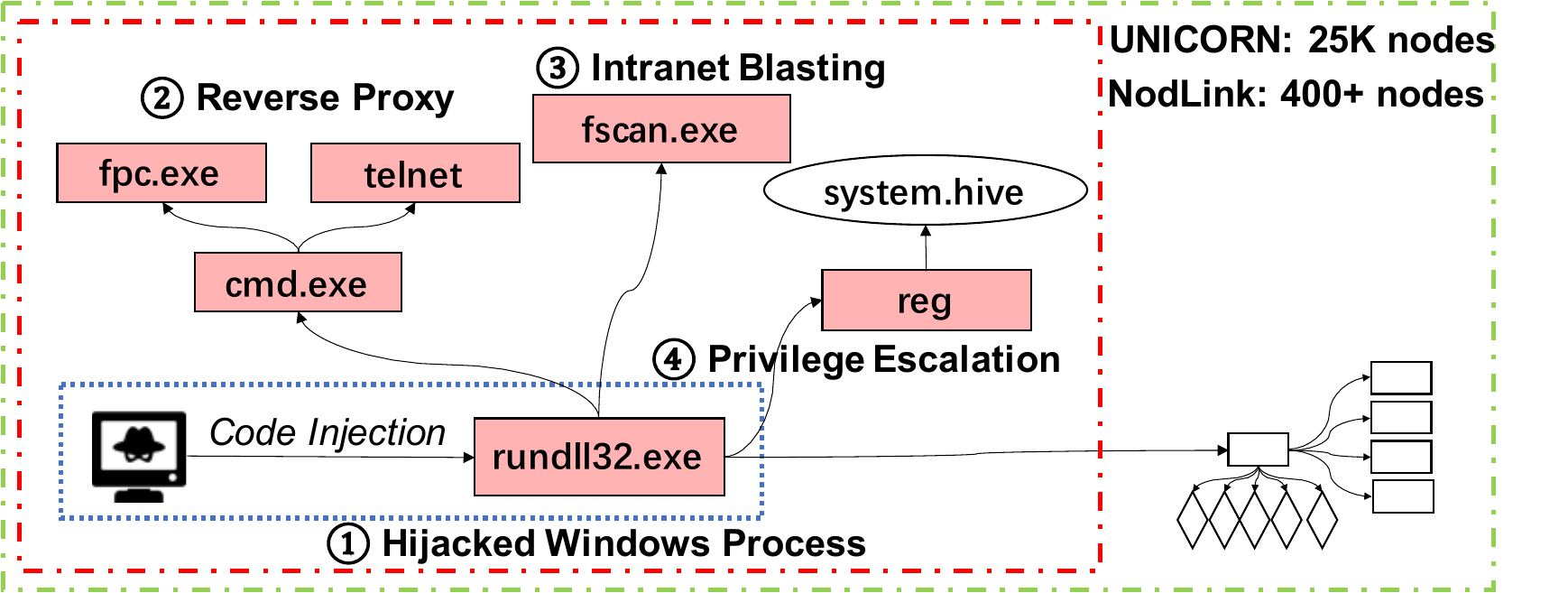}
    \caption{An attack that injects to Windows DLL Applications. Then it modifies the registry and executes the reverse proxy and intranet blasting tools. }
    \label{fig:C6}
\end{figure}

\noindent \textbf{Attack 6: }
As shown in Figure~\ref{fig:C6}, the server is attacked by malicious code memory injection into the system process ``rundll32.exe'' to set up malicious external connection. Then the malicious intranet proxy tool ``frp'' and intranet scanning tool ``fscan'' are subsequently implanted for reverse proxy and intranet blasting.  He also uses ``reg'' to modify the registry for privilege escalation.

The part boxed by the red dashed lines is the provenance graph generated by \toolname (with about 400 nodes) and the part boxed by green dashed lines is generated by UNICORN (with more than 25K nodes). The \nodeprecision of \toolname in this case is 0.20. Besides, \toolname can also identify the core attack steps, such as hijacked ``rundll32.exe'', which loads and runs 32-bit \ac{dlls}, the process for intranet blasting ``fscan.exe'', and other steps for reverse proxy and privilege escalation, as \attcandi, which are marked as red-solid boxes. In Figure~\ref{fig:C6}, we can observe that the provenance graph of UNICORN has many attack-irrelevant events that are generated by the normal behaviors of ``rundll32'' that executes the normal \ac{dlls}. 

\begin{figure}[t!]
    \centering
    \includegraphics[width=0.5\textwidth]{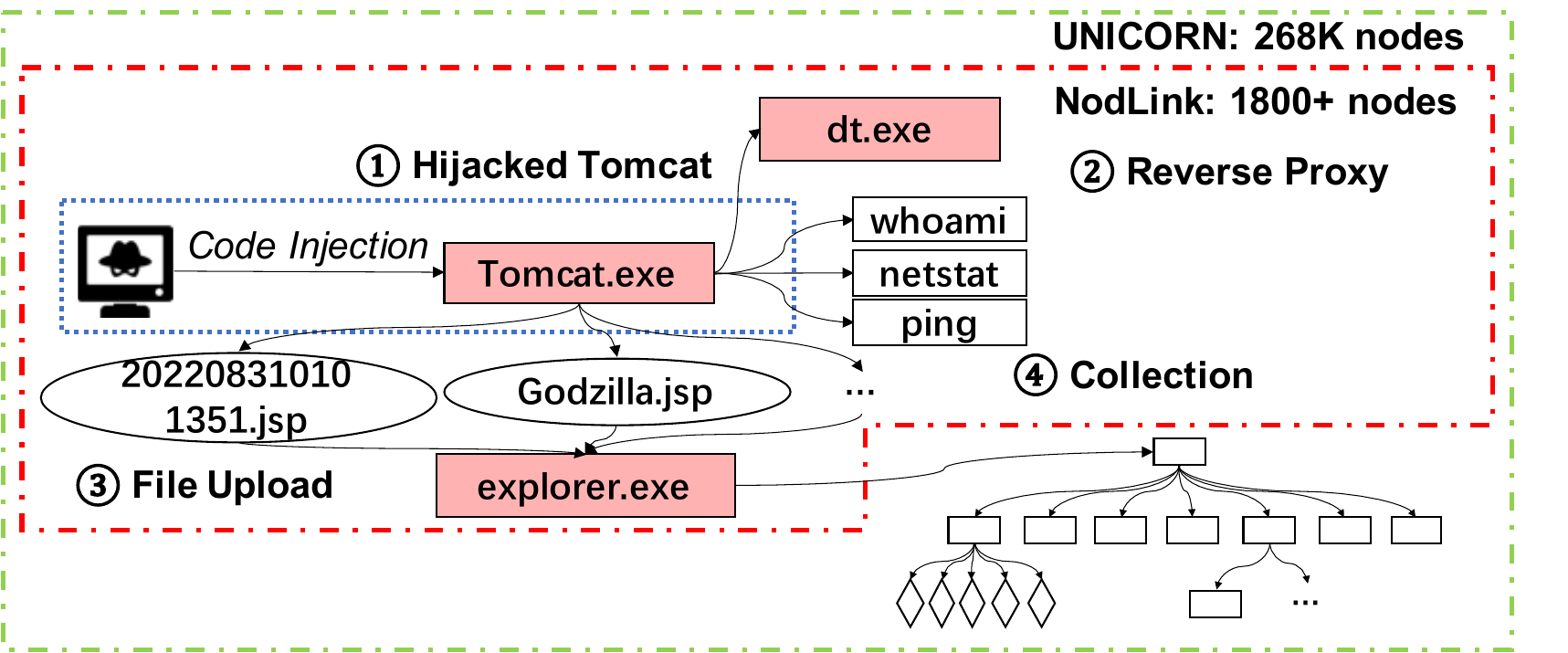}
    \caption{An attack that first injects Tomcat Server and uploads lots of malicious files, including backdoors. Then it sets up reverse proxy and collects the system information.}
    \label{fig:C7}
\end{figure}
\noindent \textbf{Attack 7: }
As shown in Figure~\ref{fig:C7}, the attacker tries to connect to the webshell backdoor of the  server ``Tomcat.exe'' and uploads a large number of files. In the temporary directory, 88 files uploaded by the attacker are found, 4 of which are Godzilla webshell and 1 is the dog-tunnel proxy tool. The rest are vulnerability test files. Then the attacker sets up ``dt.exe'' as reverse proxy.

The part boxed by the red dashed lines is the provenance graph generated by \toolname (with about 1800 nodes) and the part boxed by green dashed lines is generated by UNICORN (with more than 289K nodes). \toolname can identify the core attack steps, such as hijacked Tomcat (``Tomcat.exe''), the reverse proxy ``dt.exe'', and the file explorer that lists a large number of malicious files, as \attcandi, which are marked as red-solid boxes. In Figure~\ref{fig:C7}, we can observe that the provenance graph of UNICORN has many attack-irrelevant events that are generated by the normal behaviors of the file explorer (``explorer.exe''). 

\subsection{RQ 6: Hyperparameters}
\label{sec:parameter}
We choose the default hyperparameters using the optimal experiment results on DARPA dataset. Overall, \toolname is robust to changes in parameters because we choose to use non-parameterized learning techniques. 

\noindent \textbf{$\alpha$, $\beta$ and $\gamma$ in Hopset Construction.} 
In Hopset Construction, $\alpha$ affects node distance's impact on anomaly score. $\alpha$ values of 0.5-0.9 performed equally well on \graphacc and \nodeacc. $\beta$ should be much larger than $\gamma$ to prioritize nodes with close $AS$. We tested (100,1), (500,1), and (1000,1) for $\beta$ and $\gamma$ and got similar results on \graphacc and \nodeacc. 

\begin{figure}
    \centering
    \subfigure[$\theta$ in Hopset Construction]{
    \includegraphics[width=0.49\textwidth]{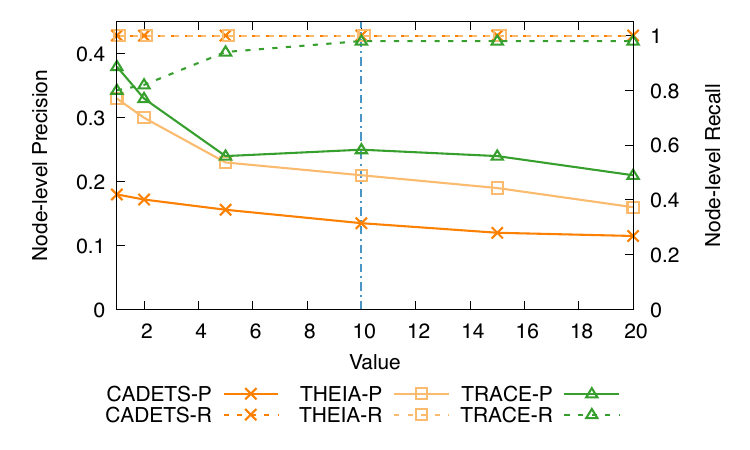}
    \label{fig:p-theta}}
    \subfigure[Decaying factor $\epsilon$]{
    \includegraphics[width=0.49\textwidth]{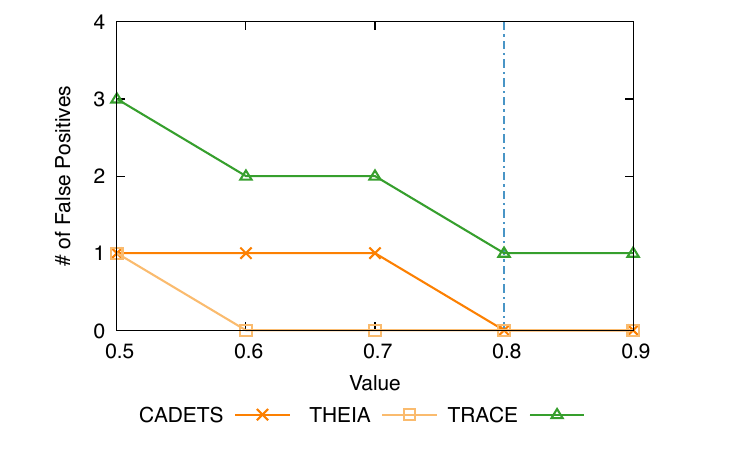}
    \label{fig:p-alpha}}
    \caption{Performance of \toolname on different parameters.}
    \label{fig:Parameters}
\end{figure}

\noindent \textbf{$\theta$ in Hopset Construction.} 
We use \nodeprecision and \noderecall to measure the impact of our tool under different $\theta$ values. $\theta$ determines the search scope for each terminal, affecting graph-level and node-level accuracy. \nodeprecision decreases and \noderecall  increases with higher $\theta$ values. Setting $\theta$ to 10 ensures complete reporting of attack scenarios while maintaining acceptable precision. The detailed result is in Figure~\ref{fig:p-theta}. 

\noindent \textbf{Decaying Factor $\epsilon$.} 
For the optimal value of the decaying factor $\epsilon$, we measured the number of false positives and true positives by varying the value of the $\epsilon$, as shown in Figure~\ref{fig:p-alpha}. In our experiment, \toolname can detect all the attacks in all the settings from 0.5-0.9, which means that the \graphrecall is 1. The results also show that,
when the decaying factor $\epsilon$ is less than 0.8, more false positives are reported. Because $\epsilon$ can affect how long a graph is kept in the cache. The smaller the value, the higher the probability that the graph will be evicted from the cache. Therefore, $\epsilon = 0.8$ in our experiments.

\section{Related Work}

King et al.~\cite{king2003backtracking} first introduced provenance graphs and backtracking, leading to numerous techniques for system auditing~\cite{190900,Wang2018FearAL, 10.1145/3427228.3427255, 10.1145/3243734.3243763, gui2020aptrace, fei2021seal,jiangauditing,liu2023adonis}, network debugging~\cite{bates2014let,wu2015automated}, and access control~\cite{park2012provenance}. Besides the online detection systems summarized in Section~\ref{sec:protocol}, SLEUTH~\cite{sleuth} is the first provenance-based online \ac{apt} detection system. Various offline post-mortem attack investigation systems~\cite{king2003backtracking,liu2018towards,hassan2019nodoze, han2021sigl,215975,216025,wang2020you,guo2011g,10.5555/1855807.1855817} exist, using heavy graph learning algorithms. However, these can delay detection, leading to financial losses~\cite{Incident-Investigation}. Compared to these systems, \toolname is the first online system that achieves high detection accuracy without losing detection granularity. There are also studies on conventional intrusion detection systems~\cite{502675,10.5555/645838.670723,10.5555/1297828.1297830,1199328,924295,10.1145/3133956.3134015,dongdistdet,jiang2023detecting}. However, these systems are orthogonal to \toolname since they cannot support \ac{apt} attack investigation. We can borrow the idea of the recent progress in conventional intrusion detection systems to improve the accuracy of anomaly identification in Section~\ref{sec:localfilter} in the future. Researchers have studied approximation algorithms for \ac{stp} that offer near-optimum solutions with polynomial running time~\cite{kou1981fast, doi:10.1137/0116001, doi:10.1137/S0895480101393155,10.5555/900518}. However, their complexity is too high for online APT detection.

\toolname is also related to conventional Host-based Intrusion Detection techniques (HIDS). 
Log2vec~\cite{10.1145/3319535.3363224} embeds the logs into vectors by graph embedding with a random walk on the heterogeneous graph. Deeplog~\cite{10.1145/3133956.3134015} models the sequence and context feature of logs using the Long-Short-Term-Memory (LSTM). These techniques are less effective in detecting APT attacks because they cannot link attack steps and reconstruct APT attack campaigns.

\section{Discussions}

\noindent\textbf{Evasion:} \toolname maintains security parity with other score-propagation-based techniques like MORSE~\cite{9152772}, PrioTracker~\cite{liu2018towards}, and NoDoze~\cite{hassan2019nodoze}, offering efficient and accurate anomaly score calculation for provenance subgraphs. Despite its robustness, two potential attacks against \toolname are identified. First, attackers could use benign decoy events to reduce an attack path's anomaly score, exploiting the decaying factor used in score propagation. However, this theoretical threat is less problematic in practice, as creating benign nodes often results in new anomalies and creates detectably anomalous path shapes. Secondly, attackers could attempt to evade detection by extending the time interval between each attack step, leading to anomalous nodes being purged from the cache. \toolname counters this through a node retrieval mechanism that reinstates evicted nodes from the disk, as described in Section~\ref{InmemoryCache}, safeguarding against long-term attack evasion.

\noindent\textbf{Robustness:}\label{sec:sectionrq7} 
\toolname is resilient to a training dataset containing minor attacks, thanks to Grubbs's test identifying and the robustness of \ac{vae}. We assessed \toolname's robustness against polluted training data using five close-world datasets, to which we added one day of attack data. Results showed \toolname's graph and node-level accuracy remained unaffected by the presence of a few attack-related data in the training set. This supports our design choice to employ \ac{vae} and non-parameterized anomaly detection techniques.

\noindent\textbf{Other OS:} \toolname supports provenance data collected from different operating systems. At present, the experimental data of \toolname is collected from Ubuntu and Windows hosts, but it also supports other systems such as openEuler.
\section{Conclusion}

Online provenance-based detection systems are preferred over post-mortem ones for detecting APT attacks in a timely manner. However, existing systems sacrifice detection granularity for accuracy due to limited resources and timeliness. We observe that existing systems fail to achieve fine-grained detection because they waste most of the limited resources on obviously benign events. To this end, we propose \toolname, an online provenance-based detection system that can achieve fine-grained detection while meeting the constraints of limited resources and timeliness. The key idea of \toolname is to model the \ac{apt} attack detection problem as an \ac{stp}, which has efficient online approximation algorithms with theoretically bounded errors. Our experiments in a production environment show that \toolname can achieve magnitudes higher detection and investigation accuracy with the same or higher throughput compared with two \ac{sota} online provenance analysis systems.

\section*{Acknowledgments}
This work was partly supported by the National Natural Science Foundation of China (62172009, 62302181, 62072046), Huawei Research Fund, CCF-Huawei Populus Grove Fund, National Key R\&D Program of China (2021YFB2701000), the Key R\&D Program of Hubei Province (2023BAB017, 2023BAB079).
Xusheng Xiao's work is partially supported by the National Science Foundation under the grant CNS-2028748.

%


\bibliographystyle{IEEEtranS}
\bibliography{references}
%


\end{document}